%% file: arxiv.tex
\documentclass{article}
\usepackage{arxiv}

\usepackage{graphicx}
\usepackage{natbib}
\usepackage{amsmath}
\usepackage{footmisc}
\usepackage{hyperref}
\usepackage{listings}
\usepackage{listings}
\usepackage[flushleft]{threeparttable}
\usepackage{dirtytalk}
\usepackage[utf8]{inputenc} % allow utf-8 input
\usepackage[T1]{fontenc}    % use 8-bit T1 fonts
\usepackage{url}            % simple URL typesetting

\input{listingsconfig.tex}

\usepackage[many]{tcolorbox}    	% for COLORED box
\definecolor{boxsummarybackground}{HTML}{efefef}
\newtcolorbox{boxsummary}{
    colback = boxsummarybackground, % background color
    boxrule = 0pt  % no borders
}

\title{Can a domain-specific language improve program structure comprehension of data pipelines? A mixed-methods study.}

\newif\ifuniqueAffiliation
\uniqueAffiliationtrue

\ifuniqueAffiliation % Standard variant of author block
\author{ \href{https://orcid.org/0000-0002-4236-2689}{\includegraphics[scale=0.06]{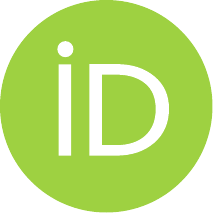}\hspace{1mm}Philip~Heltweg} \\
	Friedrich-Alexander-Universität Erlangen-Nürnberg\\
	Erlangen, Germany \\
	\texttt{philip@heltweg.org} \\
	\And
	\href{https://orcid.org/0000-0001-9060-7938}{\includegraphics[scale=0.06]{orcid.pdf}\hspace{1mm}Georg-Daniel~Schwarz} \\
	Friedrich-Alexander-Universität Erlangen-Nürnberg\\
	Erlangen, Germany \\
	\texttt{georg.schwarz@fau.de} \\
        \And
        \href{https://orcid.org/0000-0002-8139-5600}{\includegraphics[scale=0.06]{orcid.pdf}\hspace{1mm}Dirk~Riehle} \\
	Friedrich-Alexander-Universität Erlangen-Nürnberg\\
	Erlangen, Germany \\
	\texttt{dirk@riehle.org} \\
}
\else
\usepackage{authblk}

\newbox{\orcid}\sbox{\orcid}{\includegraphics[scale=0.06]{orcid.pdf}} 
\author[1]{%
	\href{https://orcid.org/0000-0000-0000-0000}{\usebox{\orcid}\hspace{1mm}David S.~Hippocampus\thanks{\texttt{hippo@cs.cranberry-lemon.edu}}}%
}
\author[1,2]{%
	\href{https://orcid.org/0000-0000-0000-0000}{\usebox{\orcid}\hspace{1mm}Elias D.~Striatum\thanks{\texttt{stariate@ee.mount-sheikh.edu}}}%
}
\affil[1]{Department of Computer Science, Cranberry-Lemon University, Pittsburgh, PA 15213}
\affil[2]{Department of Electrical Engineering, Mount-Sheikh University, Santa Narimana, Levand}
\fi

\renewcommand{\headeright}{Preprint}
\renewcommand{\undertitle}{Preprint}
\renewcommand{\shorttitle}{Can a DSL improve program structure comprehension of data pipelines?}

\hypersetup{
pdftitle={Can a domain-specific language improve program structure comprehension of data pipelines? A mixed-methods study.},
pdfauthor={Philip~Heltweg, Georg-Daniel~Schwarz,Dirk~Riehle},
}

\begin{document}
\maketitle

\begin{abstract}
In many application domains, domain-specific languages can allow domain experts to contribute to collaborative projects more correctly and efficiently. To do so, they must be able to understand program structure from reading existing source code. With high-quality data becoming an increasingly important resource, the creation of data pipelines is an important application domain for domain-specific languages. 

We execute a mixed-method study consisting of a controlled experiment and a follow-up descriptive survey among the participants to understand the effects of a domain-specific language on bottom-up program understanding and generate hypotheses for future research.

During the experiment, participants need the same time to solve program structure comprehension tasks, but are significantly more correct when using the domain-specific language. In the descriptive survey, participants describe reasons related to the programming language itself, such as a better pipeline overview, more enforced code structure, and a closer alignment to the mental model of a data pipeline. In addition, human factors such as less required programming experience and the ability to reuse experience from other data engineering tools are discussed.

Based on these results, domain-specific languages are a promising tool for creating data pipelines that can increase correct understanding of program structure and lower barriers to entry for domain experts. Open questions exist to make more informed implementation decisions for domain-specific languages for data pipelines in the future.
\end{abstract}

% keywords can be removed

\include{introduction}
\include{related-work}
\include{methods}
\include{results}
\include{discussion}
\include{limitations}
\include{conclusion}

\include{acknowledgements}
\include{data-availibility-statement}

\include{conflict-of-interest}

\bibliographystyle{apalike}       % basic style, author-year citations
\bibliography{arxiv.bib}   % name your BibTeX data base

\include{appendix}

\end{document}

%% file: listingsconfig.tex
\lstdefinelanguage{Jayvee}{
  morekeywords=[1]{block, oftype, pipeline, composite, blocktype, property, input, output, constraint, valuetype, on},
  morekeywords=[2]{true, false, text, decimal, integer, None, Sheet, value},
  morekeywords=[3]{},
  sensitive,
  morecomment=[s]{/*}{*/},
  morecomment=[l]//,
  morecomment=[s]{/**}{*/}, % JavaDoc style comments
  morestring=[b]',
  morestring=[b]"
}[keywords, comments, strings]

%% file: introduction.tex
\section{Introduction}
\label{sec:introduction}
Domain-specific languages (DSLs) can be a useful alternative to general-purpose programming languages (GPLs) in many application domains. By focussing on one domain, they can have a reduced scope and re-use glossary and concepts from the application domain, making them easier to learn and more efficient to program for domain experts \citep{Kosar2018-ck, Johanson2017-sb}. However, because DSLs are a specialized tool, they have to be carefully evaluated to determine whether they provide enough benefits to make their adoption a good choice.

When working on non-trivial software applications, developers must first understand the program structure from source code. Only then can they make changes to extend existing implementations or fix bugs. Program comprehension, in general, is estimated to be the dominant activity while programming, with more than 50\% of time spent \citep{Roberto-Minelli2015-yo, Xia2018-hl}. Therefore, the effects of a DSL on program structure comprehension are essential for the usefulness of a DSL in an application domain.

The evaluation of DSLs generally has to be domain-specific \citep{Kosar2018-ck}. Increasingly, high-quality data, and with it data engineering, is of large importance in industry because many innovative apps and AI applications rely on access to data. Sources for data sets vary from company internal data to open data, with open data mainly published by governments but also by some private entities.

Depending on the type of data, creating an automated data pipeline is a major part of data engineering. An example is regularly changing data, such as schedules released as open transport data, that should be ingested and improved automatically with updated releases.

In complex domains, data-engineers must collaborate with subject-matter experts to understand the meaning of data. A common challenge during these collaborations is that subject-matter experts lack programming experience, which complicates it to find a shared collaboration artifact with professional programmers \citep{Heltweg2023-ps}.

Domain-specific languages can be a useful middle-ground, that enables subject-matter experts to contribute directly to the creation of data pipelines, as previously shown in other domains \citep{Johanson2017-sb, Lopes2021-ol}.

DSLs can be grounded in the formal and informal glossary of domain experts, such as sketches \citep{Wile2004-rr}. A common mental model for a data pipeline is a graph of processing steps connected by pipes, known from visual programming. A DSL can provide an explicit syntax and semantics to express this data pipeline structure with the pipes and filters architecture.

In previous explorative work, we found using a domain-specific language based on this architecture had positive effects on speed, quality of the solution and perceived difficulty when solving data engineering exercises on real life open data sets \citep{Heltweg2025-uc}.

Building on this high-level validation, we aim to understand how domain-specific languages contribute to improved performance by subject-matter experts and what language features are important in more detail. To do so, we conduct a series of empirical evaluations using quantitative and qualitative methods. Previous research shows that programming language research lacks empirical studies, instead focusing on solution proposals \citep{Do_Nascimento2012-cm}. However, empirical user studies to evaluate usability are essential tools that can lead to insights that would not have been gained otherwise \citep{Buse2011-ra, Barisic2018-kt}.

When contributing to a collaborative data engineering project, the first thing a subject-matter expert will need to do is read and understand the intention behind data pipeline source code. To start, we therefore focus on bottom-up program comprehension, the process of inferring the intentions behind an implementation from reading source code \citep{Wyrich2023-ji}; we do so in the domain of building data pipelines by non-professional programmers (subject-matter experts).

In the context of this mixed-methods study, we compare data pipelines implemented in a DSL using an explicit pipes and filters architecture (Jayvee) to imperative scripts in a GPL with libraries for data engineering (Python with Pandas). We performed an initial controlled experiment to gather quantitative data on task performance in terms of time and correctness. In a follow-up survey, we look for causal influences for the experiment outcomes..

With the results, we answer the following research questions:

\textbf{Research Question 1:} Do data pipelines implemented in Jayvee change bottom-up program structure comprehension compared to Python/Pandas for non-professional programmers...

\textbf{a:} regarding speed?

\textbf{b:} regarding correctness?

\textbf{c:} regarding the perceived difficulty?

\textbf{Research Question 2:} What reasons exist for effects on program comprehension for data pipelines implemented in Jayvee compared to Python/Pandas for non-professional programmers?

In this article, we contribute:

\begin{enumerate}
    \item A mixed-methods approach, combining a controlled experiment with a descriptive survey, to evaluate the effects of DSLs in the domain of data pipeline modelling.
    \item Quantitative data, based on a controlled experiment, on how strongly the use of a DSL in the domain of data pipeline modelling can influence pipeline structure understanding, contributing to the growing literature on domain-specific languages and motivating their use in data engineering.
    \item Explanations for these effects from participant surveys to guide future practitioners or researchers that implement DSLs for data engineering.
\end{enumerate}

%% file: related-work.tex
\section{Related Work}
\label{sec:related-work}

Empirical research into the effects of domain-specific languages has been performed across multiple domains. Kosar et al. have used controlled experiments to compare DSLs with GPLs and libraries. Initially, in the context of GUI programming, they compared the DSL XAML with C\# Forms, with XAML performing better for answering questions on provided source code \citep{Kosar2010-cp}.

With a similar approach, \citet{Kosar2012-ap} extended the insights to the domains of feature diagrams and graphical descriptions, again comparing a DSL with a GPL and an appropriate library.
While the previous experiments were performed on paper, a replication study in \citet{Kosar2018-ck} allowed the use of IDEs. In all studies, participants performed more accurate and efficient in program comprehension tasks using a DSL than a GPL with libraries.

Similar to our work, Kosar et al. have evaluated the use of DSLs for different domains using experiments and note that because DSLs are domain-specific, they must be evaluated for each domain. Our goal is to extend their work with the domain of creating data pipelines for data engineering. In addition to a purely quantitative comparison of performance, we also provide qualitative insights into potential reasons for different performance.

Other DSLs with similar structure, either for data pipelines or using blocks, have been proposed. \citet{Cingolani2015-tr} present an external DSL for the creation of data pipelines in the domain of biological data called BigDataScript. They similarly plan to support subject-matter experts, but do so by replicating script-style programming and abstracting from the underlying architecture. In contrast to our work, they demonstrate the independence towards architecture, robustness, and scalability of the language implementation technically but do not evaluate it empirically.

PACE is an external DSL for continuous integration pipelines with a block structure that compiles to JSON, presented in \citet{Fonseca2020-tv}. In a controlled experiment, participants are tasked with pipeline creation and extension while thinking aloud, comparing PACE with their previous system of manually creating JSON configs with the results showing an improvement using PACE. We use a similar mixed-methods research design, however, in a very different context (understanding data pipelines by non-professional developers instead of creation of CI pipelines in an industrial setting).

In their PhD thesis, \citet{Misale2017-wo} designed and developed PiCo, a DSL based on pipes and the data flow computational model. They demonstrate the capability of their design and evaluate the performance of the implementation using case studies and experiments with Flink and Spark. In comparison, our work provides an empirical evaluation of code comprehension instead.

As with our study, students are commonly used as participants in controlled experiments, which can provide useful data if their use as a proxy for a specific type of developer is appropriate \citep{Falessi2018-so}.

Lopes et al. compared a text-based DSL with a graphical tool in a different domain (entity-relationship modeling) with students \citep{Lopes2021-ol}. Their results are aligned with ours, showing that a textual approach using a DSL is possible with a slight advantage in quality but no difference in effort.
A similar controlled experiment on readability (speed and correctness) of type inference rules shown in a DSL or Java implementation is described in \citet{Klanten2024-pf}. The authors point out that research into programming language design lacks empirical studies, a research gap our work contributes to reducing.

Hoffmann et al. evaluated Athos, a DSL that targets subject-matter experts in the domain of vehicle routing and traffic simulation, compared to JSpirit (Java with libraries) \citep{Hoffmann2022-ab}. As with the previous studies, the DSL improved efficiency. In addition, participants reported improved user satisfaction when using Athos. Even with the planned end users being subject-matter experts, the authors rely on students as proxies for subject-matter experts.

Similar to these studies, our work uses students as participants because we consider them a good approximation for practitioners that had first programming experiences but are not professional developers (such as subject-matter experts that have to do data engineering).

Empirical evaluations of DSLs with subject-matter experts are rare. An example is \citet{Johanson2017-sb} in which ecologists use the Sprat Ecosystem DSL and the GPL C++ to solve program comprehension tasks related to ecosystem simulations. Participants are subject-matter experts from a non-technical domain (marine science) with only moderate previous programming experience. Time to task completion and correctness were measured, with the tasks being solved in less time and with higher correctness using the DSL. The context of our research are also subject-matter experts and not technical users. We extend the insights gathered in this study by investigating a different domain (the creation of data pipelines).

%% file: methods.tex
\section{Methods}
\label{sec:methods}
\begin{figure}[!ht]
    \includegraphics[width=\textwidth]{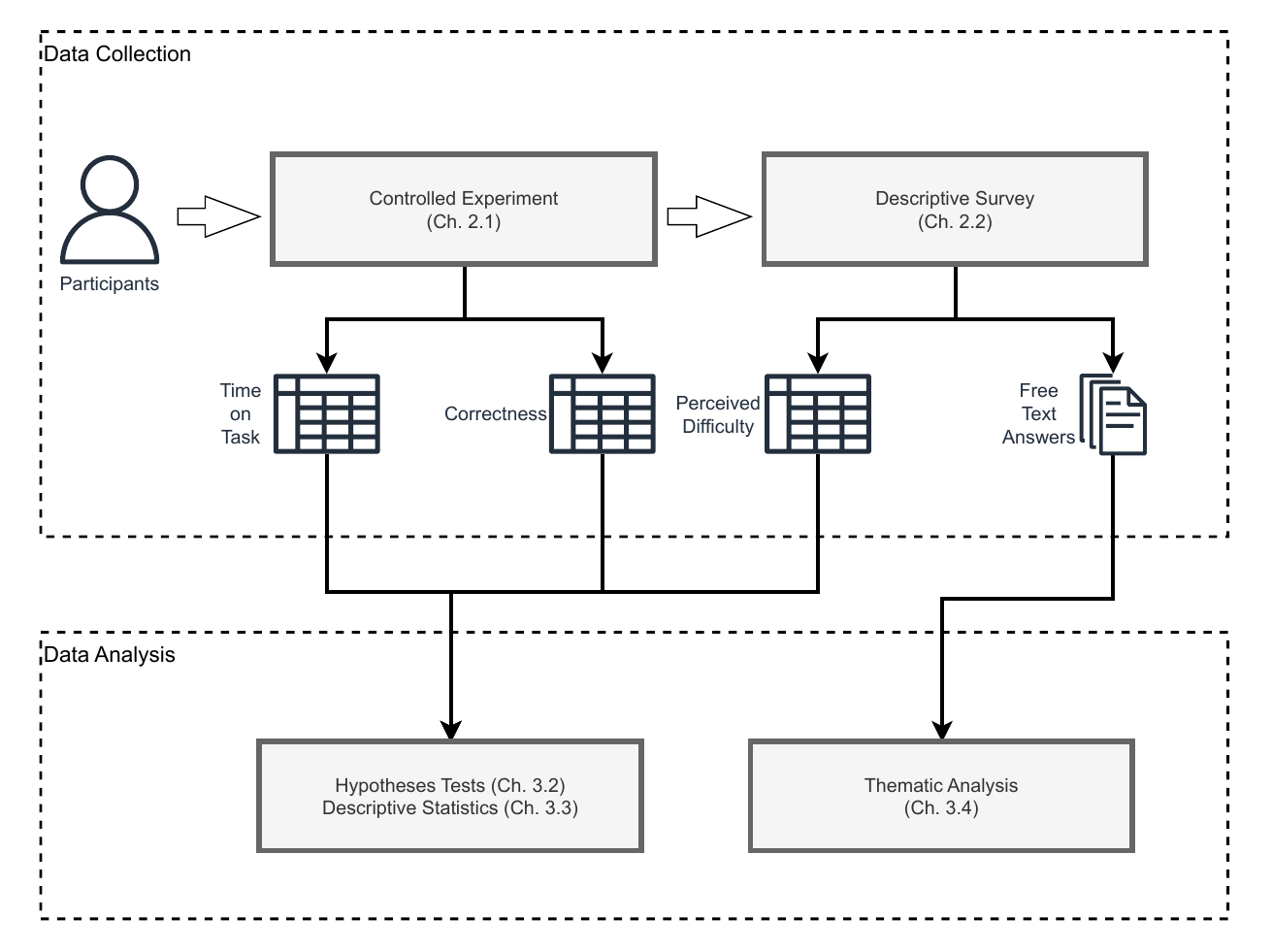}
    \caption{Overview of the mixed method research design, split into data collection and data analysis.}
    \label{fig:research-design}
\end{figure}

We used a mixed method research design \citep{Johnson2007-kr}, combining quantitative data from a controlled experiment according to \citet{Ko2015-ti} and a descriptive survey according to \citet{Kitchenham2008-ki} with qualitative data from free-text responses to the same survey. We chose thematic analysis according to \citet{Braun2012-cm} to extract common themes from the survey responses. An overview of the complete research design is shown in \autoref{fig:research-design}.

The combination of these methods allows us to validate our hypotheses in a rigorous manner and uncover potential causal relationships that strengthen the insights and enable us to generate further hypotheses to test in future work. Additionally, the qualitative responses also touch other topics of program comprehension in addition to program structure, allowing us to describe a wider diversity of effects regarding \textit{RQ2}.

\subsection{Jayvee, a Domain-Specific Language for Data Pipelines}
\label{subsec:jayvee}
Jayvee is a DSL for data engineering following the well-known pipes and filters architecture described in \citet{Garlan1993-cz} and \citet{Shaw1995-hz}. The language is designed to align as closely as possible with the mental model of data pipelines as directed acyclical graphs of processing steps, thereby making it easier for subject-matter experts to use than traditional GPLs.

The main elements of a Jayvee pipeline model is a top level \texttt{pipeline}, consisting of multiple \texttt{block}s, each representing a processing step. The inputs and outputs of these blocks are connected using \texttt{pipe}s. Blocks have an \texttt{oftype} relationship with \texttt{blocktype}s, which defines the input and output types of the block as well as its properties that can be configured.

Jayvee is an external DSL that is not embedded in a host programming language but has its own syntax and semantics. The syntax is implemented using a context free grammar language provided by Langium\footnote{\url{https://langium.org/}} while a TypeScript based interpreter acts as a reference implementation for the language semantics.

\autoref{fig:jayvee-example} shows an example of a data pipeline implemented in Jayvee. The pipeline consists of three \texttt{block}s, each performing a step in the data processing. At the top of the pipeline definition (line 2-4), the pipeline structure is defined by connecting the blocks using the \texttt{pipe} syntax \texttt{->}.

\input{jayvee-example.tex}

Jayvee includes more advanced concepts such as user-defined value types and a standard library of prebuilt, domain-specific blocks. The language is open source and available on GitHub\footnote{\url{https://github.com/jvalue/jayvee}}, additional documentation is hosted at \url{https://jvalue.github.io/jayvee}.

\subsection{Controlled Experiment}
\label{subsec:controlled-experiment}
We follow the guidelines on reporting experiments described in \citet{Wohlin2012-ze}, originally by \citet{Jedlitschka2005-mu}. We first provide informal information about research goals and the context of the experiment, and then report details of the experimental design. The experiment execution and resulting data is reported in \autoref{sec:results}.

We followed the Goal/Question/Metric template to define the research objective of the controlled experiment \citep{Wohlin2012-ze, Basili1988-al}:

\begin{enumerate}
    \item Analyze \textit{a DSL and a GPL with a specific data engineering library}
    \item for the purpose of \textit{their effect on bottom-up program structure comprehension for data pipelines}
    \item with respect to \textit{speed and correctness}
    \item from the point of view of \textit{researchers}
    \item in the context of \textit{a university course with masters level students learning data science (as proxies for non-professional programmers).}
\end{enumerate}

Our goal was to understand the influence of a DSL on professionals of non-programming disciplines that work with data as part of their jobs. Some examples include data scientists or subject-matter experts, e.g., in biology, that analyze data. Representatives from this population have base programming skills from working with data, but are not professional software engineers. 

The experiment was conducted with student participants in person, over two days in computer labs provided by the university. During the experiment, participants solved two program structure understanding tasks by reading source code of a pipeline and recreating the data pipeline structure afterward.

\begin{figure}[!ht]
    \includegraphics[width=\textwidth]{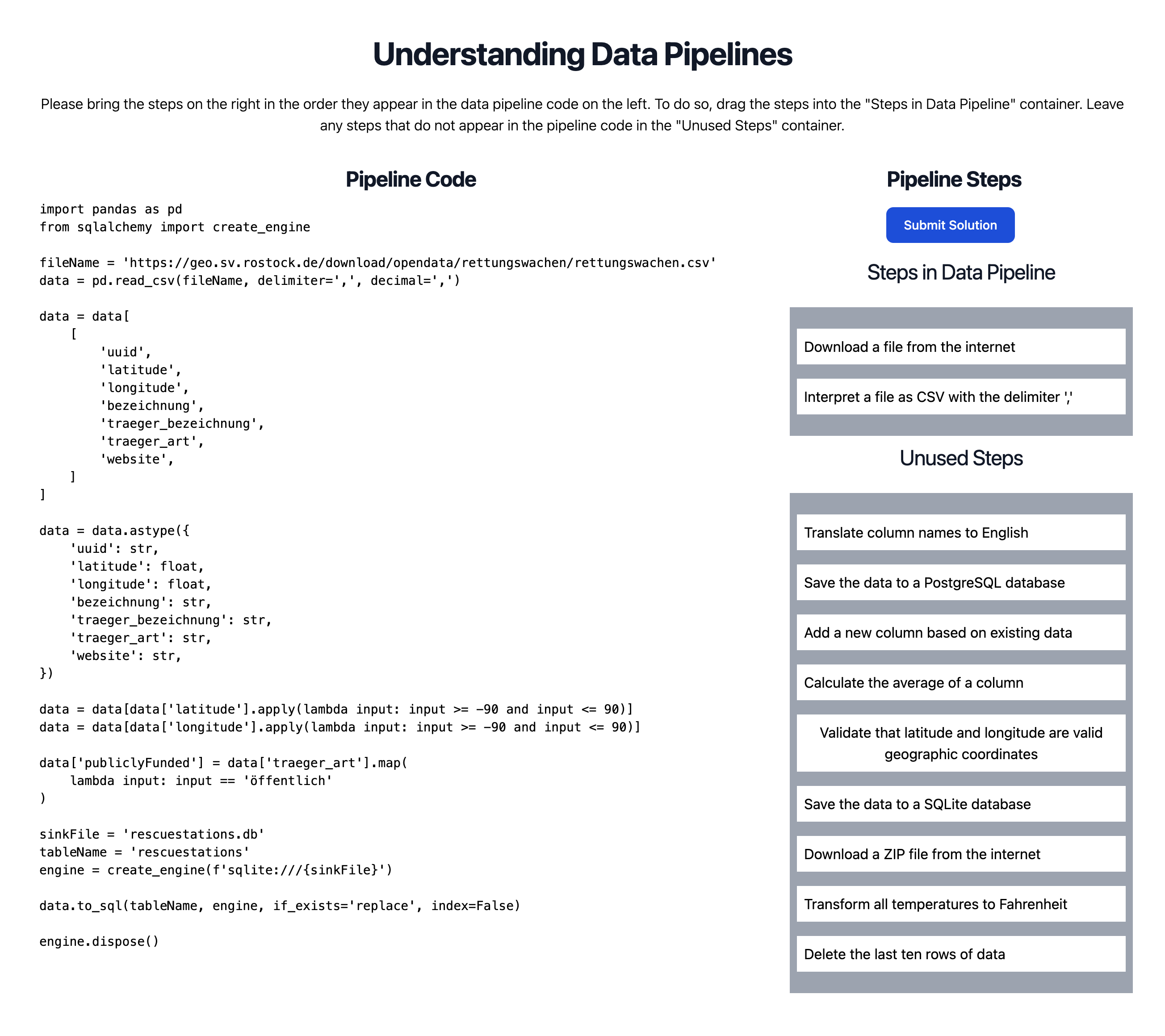}
    \caption{The experiment tool during task 2 in Python/Pandas. Pipeline source code is shown on the left, the recreation using ordered steps on the right.}
    \label{fig:experiment-screen}
\end{figure}

We use a concrete example task as an overview before describing the experiment design in detail in the following sections. \autoref{fig:experiment-screen} is a screenshot of a task view in the web-based experiment tool the participants used. On the left-hand side, under \textit{Pipeline Code} the source code of a data pipeline is shown. This data pipeline was implemented either in Jayvee or Python/Pandas, depending on the treatment group. On the right-hand side, under \textit{Pipeline Steps}, participants had to recreate the data pipeline structure by dragging steps from the list of \textit{Unused Steps} into the \textit{Steps in Data Pipeline} and bringing them into the correct order. Once they were satisfied with their solution, they could submit it using the \textit{Submit Solution} button and attempt the next task.

\subsubsection{Goals, hypotheses, and variables}
\label{subsec:goals-hypotheses-variables}
We defined one independent variable, the programming language $PL$ used to implement a data pipeline, either Jayvee ($JV$) or Python/Pandas ($PY$).

From the research objectives, we chose time to task completion and correctness as dependent variables. The combination of time and correctness is the most common for comprehension tasks \citep{Wyrich2023-ji}.

\textbf{Time to completion} describes the time between seeing the source code and submitting a solution. At the start of each task, the source code of the data pipeline was hidden so participants could read the available steps they had to categorize and order. We started the time measurement once participants revealed the source code by pressing a button. The time is directly measured in milliseconds by the experiment software and defined as follows:

\begin{equation}
    time(PL): \textrm{time of submission for task in PL} - \textrm{time of source code reveal}
\end{equation}

\textbf{Correctness} is an indirect variable that is calculated from the submitted solution by the participant.

For each task, $n$ potential steps are available for participants to choose from. A subset of these available steps is present in the pipeline, in a specific order. Using the drag and drop interface, participants can categorize steps into \textit{Steps in Data Pipeline} or used steps and \textit{Unused Steps} and decide on an order of steps inside these categories. 

To define the correctness of a solution $S$, we consider two dimensions: Has the participant correctly understood which steps exist in the pipeline source code and have they understood the order in which they are executed?

Regarding \textit{existence}, we count the number of steps that have been categorized correctly, either steps that exist in the pipeline and have been categorized as used or steps that do not exist in the pipeline and have been categorized as unused. Because each step can only be assigned to one category, the maximum number of correctly categorized steps is equal to $n$.

Regarding \textit{order}, we count the number of swaps needed to bring the steps that the participants categorized as used into the correct order, ignoring incorrectly categorized steps. As a sorting algorithm, we chose selection sort because it requires the minimal number of swaps to sort. We chose to handle the order of steps by sorting instead of comparing with a reference solution because a small error in ordering could mean all following steps are also in the wrong position, which would lead to a large penalty for small errors. In contrast, if the correct order can be reestablished with few swaps, the penalty is more appropriate.

In combination, we can define correctness as follows:

\begin{multline}
    correctness(PL):\\
    \frac{\textrm{\#correctly categorized steps} - \textrm{\#swaps needed for correct order}}{n}
\end{multline}

Based on this definition, correctness is a numeric value $[0, 1]$. For example, if a task has 10 available steps, and the participant categorized 9 of them correctly and in the right order, the correctness would be $0.9$. If a participant instead categorized all steps correctly, but two swaps were needed to bring the used ones into the correct order, the correctness would be $0.8$.

\textbf{Hypotheses} were defined based on the goal to describe effects on speed and correctness.

For \textit{speed}, we defined $H_{0, 1}$ as \textit{"Non-professional programmers need the same time to understand the structure of a data pipeline model when implemented in Jayvee compared to Python/Pandas."} with the alternative hypothesis $H_{1, 1}$, \textit{"Non-professional programmers do not need the same time to understand the structure of a data pipeline model when implemented in Jayvee compared to Python/Pandas."}. More formally:

\begin{equation}
    \begin{aligned}
        H_{0, 1}: time(JV) = time(PY) \\
        H_{1, 1}: time(JV) \neq time(PY)
    \end{aligned}
\end{equation}

Regarding \textit{correctness}, we defined $H_{0, 2}$ as \textit{"Non-professional programmers understand the structure of a data pipeline model equally correct when implemented in Jayvee compared to Python/Pandas."} with the alternative hypothesis $H_{1, 2}$, \textit{"Non-professional programmers can understand the structure of a data pipeline model not equally correct when implemented in Jayvee compared to Python/Pandas."}. More formally:

\begin{equation}
    \begin{aligned}
        H_{0, 2}: correctness(JV) = correctness(PY) \\
        H_{1, 2}: correctness(JV) \neq correctness(PY)
    \end{aligned}
\end{equation}

\subsubsection{Experiment Design}
We chose a factorial crossover design according to \citet{Vegas2016-dr} which is a within-subjects design in which each participant is assigned to every treatment exactly once. Crossover designs are well understood and commonly used for software engineering experiments \citep{Wyrich2023-ji}.

The participants completed two tasks reading a data pipeline, implemented in either Jayvee or Python/Pandas and recreating it using a drag and drop interface. We defined two periods (solving task 1 and task 2) and two sequences AB and BA, see \autoref{tab:experiment-design}. Participants were randomly assigned to either sequence without experimenter input, based on a call to JavaScript \lstinline|Math.random| when they opened the experiment tool. One experiment session included both periods.

\begin{table}[!ht]
    \centering
    \caption{Factorial crossover design of the controlled experiment according to \citet{Vegas2016-dr}}
    \label{tab:experiment-design}
    \begin{tabular}{lll}
    \hline\noalign{\smallskip}
             & \multicolumn{2}{c}{Period}  \\
    Sequence & Task 1 & Task 2  \\
    \noalign{\smallskip}\hline\noalign{\smallskip}
    AB & Jayvee & Python/Pandas \\
    BA & Python/Pandas & Jayvee \\
    \noalign{\smallskip}\hline
    \end{tabular}
\end{table}

\subsubsection{Participants}
The experiment was executed during a masters level course on data engineering and working with open data, offered to students largely studying data science and artificial intelligence as well as some students from computer science and information systems. Because the participants are students and the vast majority of them study degree programs that mainly work with data in a theoretical fashion rather than teach software engineering, they have limited experience programming but have worked on data engineering before. We considered this population an appropriate proxy, as discussed in \citet{Falessi2018-so}, for data practitioners that have some experience with programming but are not professional software engineers. 

During the course, students were introduced to Jayvee in two lectures and were encouraged to use Python with Pandas for an individual data science project. The course requires the completion of five data engineering exercises in Jayvee and Python/Pandas, with students switching languages after each exercise. In all lectures that referenced programming challenges, we used examples in Jayvee and Python/Pandas. While we mentioned alternative libraries, we always used Python in combination with Pandas during the module.

We employed convenience sampling from this population by offered students to voluntarily participate in the experiment in place of completing the third homework exercise. Doing so would count as passing the exercise, and enter them into a raffle to win two gift cards of EUR 20 each. If they chose to complete the exercise as normal, they experienced no negative effects, e.g., their grade was unaffected.

\subsubsection{Objects, Instrumentation, and Data Collection Procedure}

Participants were asked to complete two bottom-up code comprehension tasks in which they had to read the provided source code of a data pipeline and recreate the structure using a drag and drop interface. They completed one task reading a pipeline implemented in Jayvee and one with a pipeline implemented in Python/Pandas, depending on their sequence assignment. Both tasks used a web-based experiment tool (see \autoref{fig:experiment-screen} for a task screen example) and followed the same sequence:

\begin{enumerate}
    \item Participants were shown the available steps, categorized as unused, while the pipeline source code was hidden.
    \item After reading the available steps, participants reveal the pipeline source code using a button press (time measurement starts).
    \item Participants drag and drop steps into the \textit{Steps in Data Pipeline} category and bring them in the correct order as they understand the pipeline.
    \item When they are satisfied with their solution, participants click on "Submit Solution" (time measurement stops).
    \item They are taken to a pause screen where they can start the next task whenever they feel ready.
\end{enumerate}

In addition to time measurements, the experiment tool automatically saved the submitted solution so that correctness could be calculated in the analysis phase. After both tasks, the participants were asked to complete a follow-up survey. The exact version of the tool used by participants can be found online \footref{fn:replication-package}.

Both languages were shown as text without syntax highlighting. Two researchers were in the room for every experiment run to monitor the screens of participants and ensure silence. This made sure that participants did not interact with each other or search for solutions on the internet.

For the tasks, we implemented equivalent data pipelines in Jayvee 0.1.0 and Python 3.11 with Pandas 2.0, based on real open data sources.

\begin{enumerate}
    \item Task 1 is a pipeline that downloads a ZIP-file, extracts it and selects a file as CSV. It then translates some columns names to English, selects a subset of columns and saves the data to a SQLite-database.
    \item Task 2 is a pipeline that downloads a file, interprets it as CSV and validates that data in one column are geographic coordinates between -90 and 90. It adds a new column with boolean data, based on another column. Finally, it saves the data to SQLite.
\end{enumerate}

We aligned the code structure as much as possible by implementing each step similarly in Jayvee and script-style Python/Pandas. As an example, \autoref{fig:task-comparison-extract} compares the source code to extract a CSV file for task 1 in both languages. The example shows the more verbose syntax of Jayvee, utilizing blocks to model processing steps, compared to Python/Pandas. The appendix (\autoref{sec:appendix}) includes a further comparison of source code used in task 2 (\autoref{fig:task-comparison-schema}).

\begin{figure}[!ht]
    \input{task-comparison-extract.tex}
    \caption{Comparison of source code excerpts to extract data from a CSV source, shown for task 1 in Jayvee and Python/Pandas.}
    \label{fig:task-comparison-extract}
\end{figure}

We conducted two pilot tests to ensure the data pipeline implementations and the accompanying step descriptions are appropriate and clear. First, we shared the tasks with other researchers that were neither involved in Jayvee development nor the experiment itself. Later, we invited students from previous semesters to take the full experiment remotely while we watched their screen and asked for their feedback afterward. Based on the feedback of both pilot groups, we made minor code and wording adjustments and gained the expectation that the tasks could reasonably be completed in 10 minutes each.

We defined an experiment procedure so multiple experimenters could guide the participants through the following process:

\begin{enumerate}
    \item Read and acknowledge informed consent information.
    \item Open allowed documentation in tabs.
    \item Provide an overview about the experiment process, how tasks work and what the experiment measures. Communicate that we expect the experiment to last for roughly 30 minutes and will announce times at 10 minutes and 20 minutes.
    \item Solve an initial example task with pseudocode together with participants to familiarize them with the tool.
    \item Answer any final questions before asking the participants to start their tasks and no longer interacting with them.
    \item Participants complete both tasks and the follow-up survey.
    \item Finally, thank the participants and ask them not to share the experiment setup with other participants.
\end{enumerate}

Because we asked participants to submit their own solutions, variations can occur between participants that choose to be faster or more correct, depending on their confidence \citep{Ko2015-ti}. To reduce this effect, we asked the participants to favor correctness over speed if in doubt.

The full source code of both tasks, the experiment procedure and the informed consent handout can be found in the replication package \footnote{All links can be found in the Data Availability Statement (\autoref{sec:data-availability-statement}). \label{fn:replication-package}}.

\subsection{Descriptive Survey}
\label{subsec:descriptive-survey}
We designed a cross-sectional, descriptive survey according to \citet{Kitchenham2008-ki} to assess how participants perceived the difficulty of understanding the data pipeline from Jayvee code compared to Python/Pandas.

As part of the survey, participants completed an online questionnaire after completing the experiment, with two agreement questions \textit{How difficult was it to understand the data pipeline written in Jayvee?} and \textit{How difficult was it to understand the data pipeline written in Python?}. Answers could be given on a 5-point Likert scale. We assigned numbers from 1 (\textit{Very easy}) to 5 (\textit{Very hard}) to be able to calculate medians and defined $\mathit{difficulty}(PL)$ as the median of the answers for $JV$ and $PY$ respectively.

To answer RQ 1c: Do data pipelines implemented in Jayvee change bottom-up program structure comprehension compared to Python/Pandas for non-professional programmers \textit{regarding perceived difficulty}, we defined $H_{0, 3}$ as \textit{"Non-professional programmers do not perceive a data pipeline model as easier or harder to understand when implemented in Jayvee compared to Python/Pandas."} with the alternative hypothesis $H_{1, 3}$, \textit{"Non-professional programmers do perceive a data pipeline model as easier or harder to understand when implemented in Jayvee compared to Python/Pandas."}. More formally:

\begin{equation}
    \begin{aligned}
        H_{0, 3}: \mathit{difficulty}(JV) = \mathit{difficulty}(PY) \\
        H_{1, 3}: \mathit{difficulty}(JV) \neq \mathit{difficulty}(PY)
    \end{aligned}
\end{equation}

In addition, participants were provided free-text input fields for the questions \textit{What makes data pipelines written in Jayvee difficult/easy to understand?}, \textit{What makes data pipelines written in Python difficult/easy to understand?}, and \textit{What are the differences between Jayvee and Python that influence how easy / hard it is to understand data pipelines?}.

To analyze this qualitative data, we chose thematic analysis according to \citet{Braun2012-cm}. Because we had no preconceived theory but wanted to understand causal relationships for the experiment results, we chose an inductive approach, letting the themes emerge from the data.

During the thematic analysis, we first familiarized ourselves with the data by reading all survey responses in detail.

Afterward, we created codes from the data and constructed a codebook by grouping related codes into themes. Our goal was the creation of a codebook that is clear and themes that can be consistently understood by multiple readers. We therefore worked in iterations, with multiple authors applying the codebook to responses independently and discussing any differences in coding that emerged from unclear descriptions to improve the clarity of themes. 

For each iteration:

\begin{enumerate}
    \item We selected a subset of the responses at random
    \item The first author coded the subset of responses and afterward updated the codebook with new insights
    \item The updated codebook was shared with another author, who used the codebook to code the same subset of responses
    \item The authors met to qualitatively discuss any differences in coding and the clarity of the codebook and the codebook was updated according to the discussion
    \item The first author used the updated codebook to re-code all previous responses
\end{enumerate}

Because our goal was to explore the diversity of reasons for the effects on program comprehension, we chose theoretical saturation as a guideline to judge the maturity of our codebook, meaning no or few new insights are gained from analyzing additional data \citep{Bowen2008-xh}. We counted codes that were assigned to each survey response, as well as any codebook changes (newly created, deleted, moved or updated codes and themes). We consider theoretical saturation to be reached when codebook changes are rare (indicating that the codebook is stable), but codes are still assigned to new responses (indicating that the codebook is relevant to the topic of the response).

%% file: jayvee-example.tex
\begin{lstlisting}[
  language={Jayvee},
  caption={Data pipeline extracting CSV data and writing it to a SQLite database, written in Jayvee.},
  label={fig:jayvee-example},
]
pipeline CarDataPipeline {
  CarDataCSVExtractor
    -> CarDataInterpreter
    -> CarDataSQLiteLoader;
  
  block CarDataCSVExtractor oftype CSVExtractor
    url: "https://example.org/data.csv";
    enclosing: '"';
  }
  block CarDataInterpreter oftype TableInterpreter {
    header: true;
    columns: [
      "name" oftype text,
      // ... further assignments
    ];
  }
  block CarDataSQLiteLoader oftype SQLiteLoader {
    table: "Cars";
    file: "./cars.db";
  }
}
\end{lstlisting}

%% file: task-comparison-extract.tex
\noindent
\begin{minipage}[t][18\baselineskip][t]{.55\textwidth}
    \begin{lstlisting}[language=Jayvee]
HttpDataSource
  ->TextInterpreter
  ->CSVFileInterpreter
  //... further blocks

block HttpDataSource oftype HttpExtractor {
  url: 'https://geo.sv.rostock.de/download/opendata/rettungswachen/rettungswachen.csv';
}

block TextInterpreter oftype TextFileInterpreter {}

block CSVFileInterpreter oftype CSVInterpreter {
    delimiter: ',';
    enclosing: '"';
}
  \end{lstlisting}
\end{minipage}
\hfill
\begin{minipage}[t][18\baselineskip][t]{.39\textwidth}
  \begin{lstlisting}[language=Python]
import pandas as pd

fileName = 'https://geo.sv.rostock.de/download/opendata/rettungswachen/rettungswachen.csv'

data = pd.read_csv(fileName, delimiter=',', decimal=',')
  \end{lstlisting}
\end{minipage}

%% file: results.tex
\section{Results}
\label{sec:results}

\subsection{Participant Sample}
\label{subsec:participant-sample}
Our sample consisted of 57 volunteers from a masters level course about advanced methods of data engineering that was completed by 98 students. Students mainly came from master's degree programs in artificial intelligence, data science and computer science. At the start of the semester, we used an online survey with previously validated questions by \citet{Feigenspan2012-gv} to measure previous experience in programming generally and Python and Jayvee specifically. Median programming experience was 7 (of 10), median comparison to classmates 3, median experience in Python 4 and median experience in Jayvee 1 (all of 5). At the end of the semester, we repeated the survey and the median experience of course participants in Jayvee had increased to 3 ($n = 77$). A detailed overview of the course entry survey results can be found in \autoref{fig:experiment-participants} (\autoref{sec:appendix}).

After the course entry survey, all participants heard two lectures on Jayvee programming and solved one data engineering exercise in Jayvee as part of the training for the experiment.

Of these 57 participants, 29 were randomly assigned to sequence AB and 28 to sequence BA.

\input{results-hypotheses}

\input{results-qda}

%% file: results-hypotheses.tex
\subsection{Hypotheses tests}
We used Python 3.11 with Pingouin 0.5.5 \citep{Vallat2018} for the statistical analysis of the data. We consider tests at the standard $\alpha = .05$ to be statistically significant.

For each participant, we calculated time on task and correctness as described in \autoref{subsec:goals-hypotheses-variables}.

Initially, we performed a Shapiro-Wilk test \citep{Shapiro1965-fv} to check if the variables were distributed normally. At $\alpha = .05$, both variables were non-normal. As a result, we chose the Wilcoxon signed-rank test \citep{Wilcoxon1945-uf} as non-parametric alternative to a paired t-test because it is appropriate for paired data from the crossover experiment \citep{Wohlin2012-ze, Vegas2016-dr}.

Variable distributions are plotted as kernel-density-plots to give an overview and make it easy to see non-normality \citep{Kitchenham2017-oi}.

We report effect sizes based on the matched pairs rank-biserial correlation (RBC) as an appropriate measure of effect size for the Wilcoxon signed-rank test used for the experiment data \citep{Kerby2014-xq}. As a correlation, it is equal to the difference between proportions of favorable and unfavorable evidence, with 0 meaning no effect and positive values indicating support for $H_1$. In addition to RBC, we also report CLES as a more intuitive measure of effect size, first introduced by \citet{McGraw1992-ks}, but based on the generalization by \citet{Vargha2000-zj} to allow non-normal and ordinal data such as the survey responses on a Likert scale. We interpret CLES based on the guidelines in \citet{Vargha2000-zj} as either small ($\geq .56$), medium ($\geq .64$) or large ($\geq .71$).

\subsubsection{Hypothesis 1: Speed}

The null hypothesis we defined for speed was $H_{0, 1}$: \textit{"Non-professional programmers need the same time to understand the structure of a data pipeline model when implemented in Jayvee compared to Python/Pandas."} We therefore chose a two-sided Wilcoxon signed-rank test, with the results shown in \autoref{tab:experiment-time}.

\begin{table}[!ht]
    \centering
    \caption{Wilcoxon signed-rank test for $H_{0, 1}: time(JV) = time(PY)$}
    \label{tab:experiment-time}
    \begin{tabular}{rrrrlrrr}
    \hline\noalign{\smallskip}
    \textit{n} & $Mdn_{JV}$ & $Mdn_{PY}$ & W-val & alternative   &   p-val &   RBC &   CLES \\
    \noalign{\smallskip}\hline\noalign{\smallskip}
    57 & 252.37 & 234.23 & 750 & two-sided          &   .546 & .093 &  .52 \\
    \noalign{\smallskip}\hline
    \end{tabular}
\end{table}

We have no reason to reject the null hypothesis and accept $H_{0, 1}$: \textit{"Non-professional programmers need the same time to understand the structure of a data pipeline model when implemented in Jayvee compared to Python/Pandas."} Based on the data and the underlying distribution (see \autoref{fig:experiment-time} in \autoref{sec:appendix}), it is reasonable to conclude that the use of programming language had no significant effect on time to completion in either direction.

\subsubsection{Hypothesis 2: Correctness}
\begin{figure}[!ht]
    \includegraphics[width=\textwidth]{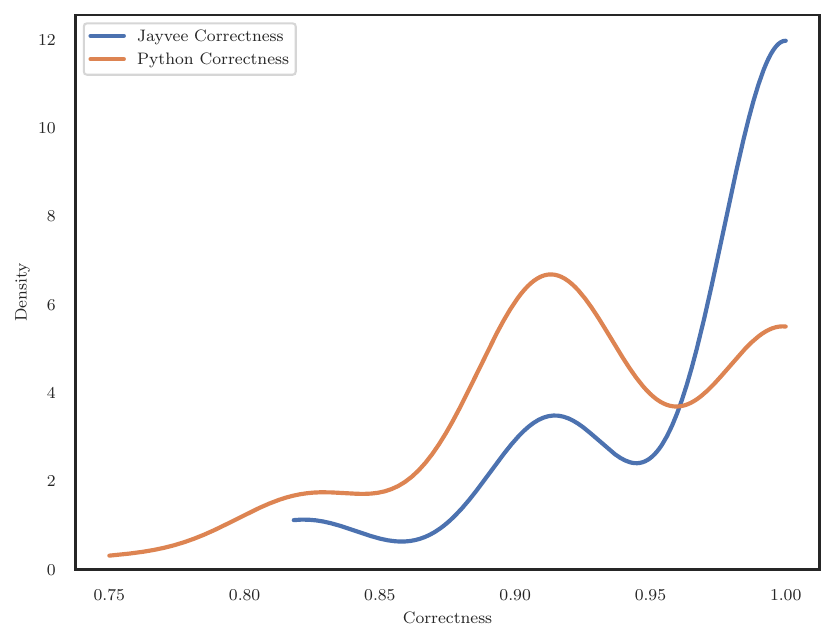}
    \caption{Kernel-density-plot of correctness of solution for Jayvee compared to Python/Pandas.}
    \label{fig:experiment-correctness}
\end{figure}

The distribution of correctness for Jayvee and Python/Pandas is plotted in \autoref{fig:experiment-correctness}.

The null hypothesis we defined for speed was $H_{0, 2}$: \textit{"Non-professional programmers understand the structure of a data pipeline model equally correct when implemented in Jayvee compared to Python/Pandas."}. We therefore chose a two-sided Wilcoxon signed-rank test, with the results shown in \autoref{tab:experiment-correctness}.

\begin{table}[!ht]
    \centering
    \caption{Wilcoxon signed-rank test for $H_{0, 2}: correctness(JV) = correctness(PY)$}
    \label{tab:experiment-correctness}
    \begin{threeparttable}
    \begin{tabular}{rrrrlrrr}
    \hline\noalign{\smallskip}
    \textit{n} & $Mdn_{JV}$ & $Mdn_{PY}$ & W-val & alternative   &   p-val &   RBC &   CLES \\
    \noalign{\smallskip}\hline\noalign{\smallskip}
    57 & 1.0 & .92 & 183 & two-sided       &   .002* & .55 &  .67 \\
    \noalign{\smallskip}\hline
    \end{tabular}
    \begin{tablenotes}
        \small
        \item * $p \leq .05$
      \end{tablenotes}
    \end{threeparttable}
\end{table}

We have reason to reject the null hypothesis and instead adopt $H_{1, 2}$: \textit{"Non-professional programmers can understand the structure of a data pipeline model not equally correct when implemented in Jayvee compared to Python/Pandas."}. The CLES indicates a medium effect size. From the distribution shown in \autoref{fig:experiment-correctness} it is clear that participants achieved significantly higher correctness when completing the experiment using Jayvee code compared to Python/Pandas. We consider this result of practical relevance because a large improvement of correctness when interpreting data pipelines will lead to significant reduced errors when working with them.

\subsection{Descriptive Survey}
The follow-up descriptive survey was filled out by 56 participants. Their impressions of difficulty for understanding the data pipelines in Jayvee and Python/Pandas were answered on a 5-point Likert scale. The exact distribution of the answers can be found in \autoref{fig:survey-difficulty} (\autoref{sec:appendix}).

After calculating medians as described in \autoref{subsec:descriptive-survey}, we again chose the non-parametric Wilcoxon signed-rank test because the data is paired and the differences in ordinal data from Likert scales can be ranked \citep{Wohlin2012-ze}. The null hypothesis we defined for speed was $H_{0, 3}$: \textit{"Non-professional programmers do not perceive a data pipeline model as easier or harder to understand when implemented in Jayvee compared to Python/Pandas."}, we therefore chose a two-sided test, with the results shown in \autoref{tab:survey-difficulty}.

\begin{table}[!ht]
    \centering
    \caption{Wilcoxon signed-rank test for perceived difficulty of using Jayvee compared to Python/Pandas, $H_{0, 3}: \mathit{difficulty}(JV) = \mathit{difficulty}(PY)$.}
    \label{tab:survey-difficulty}
    \begin{tabular}{rrrrlrrr}
    \hline\noalign{\smallskip}
    \textit{n} & $Mdn_{JV}$ & $Mdn_{PY}$ & W-val & alternative   &   p-val &   RBC &   CLES \\
    \noalign{\smallskip}\hline\noalign{\smallskip}
    56 & 2.0 & 2.0 & 380.5 & two-sided          &   .153 & -.23 &  .41 \\
    \noalign{\smallskip}\hline
    \end{tabular}
\end{table}

We have no reason to reject the null hypothesis and adopt $H_{0, 3}$: \textit{"Non-professional programmers do not perceive a data pipeline model as easier or harder to understand when implemented in Jayvee compared to Python/Pandas."}

%% file: results-qda.tex
\subsection{Qualitative Survey Responses}
\label{subsec:results-qda}

In order to identify reasons for the observed effects to answer RQ2: \textit{What reasons exist for effects on bottom-up program comprehension for data pipelines implemented in Jayvee compared to Python/Pandas for non-professional programmers?}, we used thematic analysis according to \citet{Braun2012-cm}.

To complement the quantitative data analysis of experiment results in our mixed-methods design, we collected qualitative responses to describe causal effects that might have influenced participants' task performance to open up future research directions and new hypotheses to explore. Our goal was to capture the diversity of effects that participants described rather than make additional statistical claims, so we included any relevant insight.

\begin{figure}[!ht]
    \includegraphics[width=\textwidth]{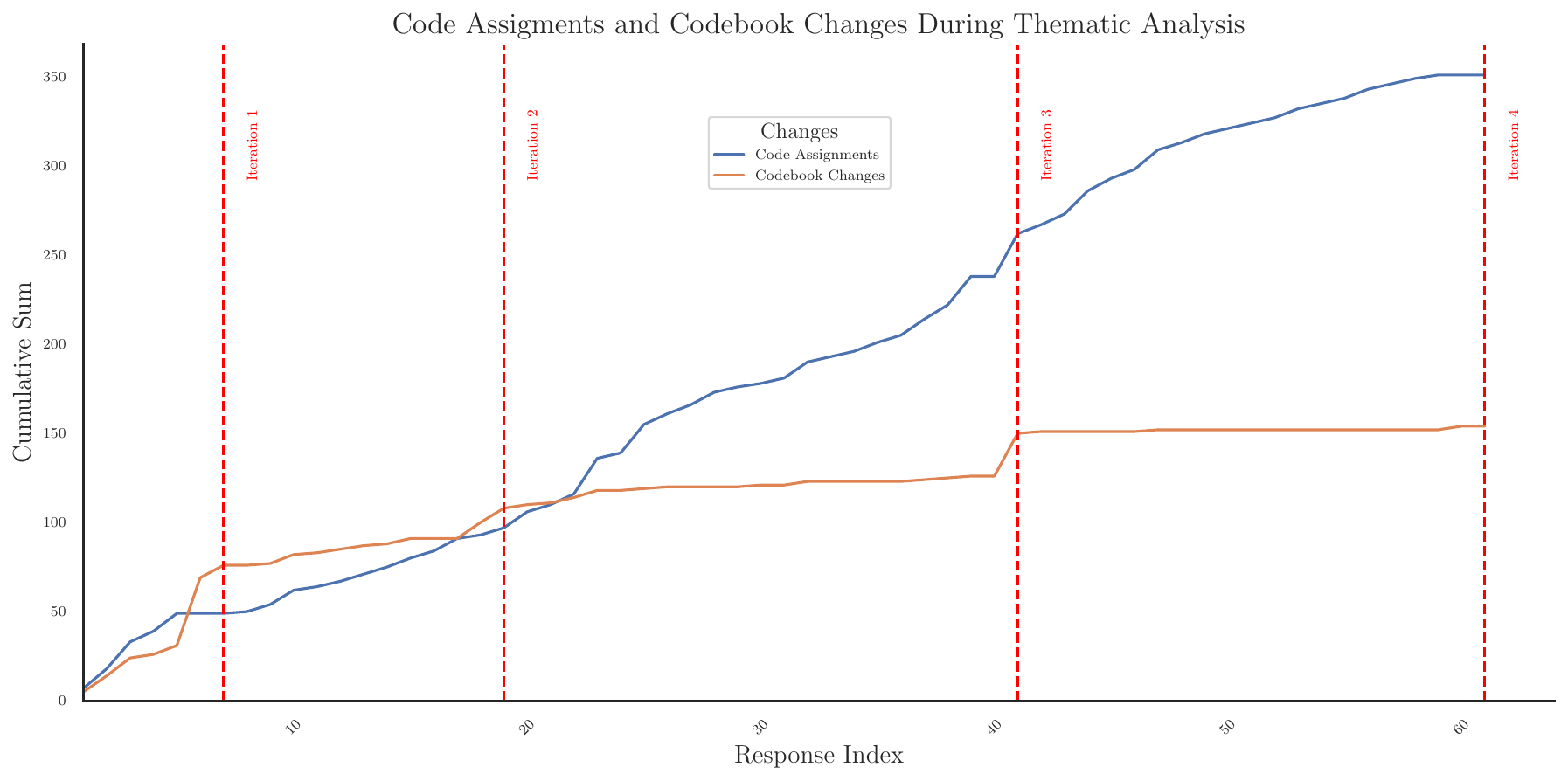}
    \caption{Code assignments compared to codebook changes during thematic analysis, showing codebook changes being rare after the third iteration, while codes were consistently applied to new responses.}
    \label{fig:qda-saturation}
\end{figure}

As described in \autoref{sec:methods}, we worked iteratively and tracked code assignments as well as codebook changes and chose theoretical saturation to judge the maturity of our theory \citep{Bowen2008-xh}. \autoref{fig:qda-saturation} shows the cumulative sum of code assignments compared to codebook changes during the thematic analysis, with every iteration highlighted by a vertical red line.

We measured inter-rater reliability using Cohen's Kappa $\kappa$ by two authors using the codebook to code new responses after every iteration. While $\kappa$ fluctuated due to the rising complexity of the codebook and the increasing number of codes, it consistently showed "substantial" agreement between the coding authors ($\kappa_{1} = .79$, $\kappa_{2} = .74$, $\kappa_{3} = .64$, $\kappa_{4} = .68$) \citep{Landis1977-ik}.

While codebook changes are frequent initially, they become much less frequent after the third iteration. Note that the high amount of codebook changes directly before the end of an iteration is due to the adaptations that are made after the qualitative discussion by the authors after coding a subset of responses. With changes being very rare during the fourth iteration, we considered theoretical saturation to be reached and are confident our codebook encapsulates the content of the survey responses well.

We present the results of our thematic analysis according to \citet{Braun2012-cm} as a collection of themes with thick descriptions. Beyond the themes that directly relate to the research questions, we also gained further insights on the role of documentation and language ecosystems. However, here we include the subset of themes that directly relate to the results from the controlled experiment. Please refer to the replication package for the full codebook with all themes and extended descriptions of codes, including additional quotes from participants \footref{fn:replication-package}.

\begin{figure}[!ht]
    \includegraphics[width=\textwidth]{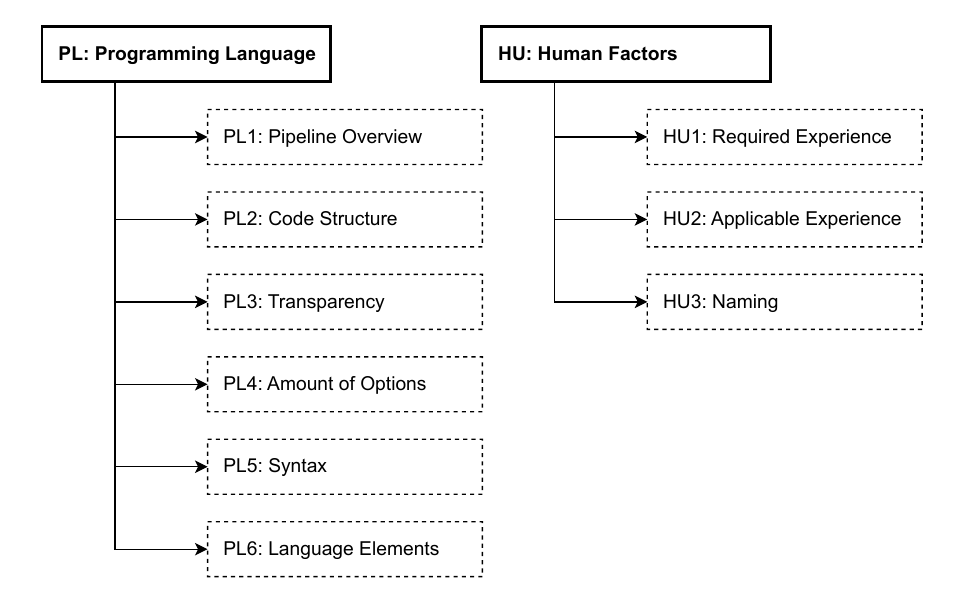}
    \caption{Overview of the codebook with two categories of themes, one related to the programming language directly and additional human factors.}
    \label{fig:codebook}
\end{figure}

\autoref{fig:codebook} shows the themes that emerged from coding, with six themes related to the programming language and three themes involving human factors.

In the rest of the chapter, we describe the themes in detail and highlight representative quotes from the surveys to give a vivid impression of the major topics in each theme.

\subsubsection{PL1: Pipeline Overview:}
\label{theme:pl1}
Jayvee splits block definitions and the wiring-up of a pipeline by connecting blocks into separate code locations (in the example Jayvee model \autoref{fig:jayvee-example}, block definitions start in line 6 while the overview is created in lines 2-4). This provides an overview of the pipeline without showing any implementation details apart from the block name.

In contrast to Jayvee with its strictly enforced structure, this overview does not always exist in procedural Python scripts that are executed from top to bottom, such as the data pipelines in the experiment. The use of Pandas does also not enforce such a structure.

A major effect of this overview is that participants \textbf{can ignore code that is not immediately needed to understand the data pipeline}. This in turn improves speed for a high-level understanding because less code has to be read as described by S18: \say{The pipeline gives a very quick overview over what happens. When the blocks are named clearly everything can be seen on one quick view.}

However, if an in-depth understanding of the implementation details is actually important to understand the data pipeline, the effect of a centralized overview on speed and understanding can potentially be negative. A few participants described a negative effect on both speed and understanding due to the additional navigation needed to read all source code. For example, S40 answered: \say{(Jayvee is difficult to understand...) due to the code structure/layout, need to go back \& forth to search for the specific function.}

The centralized overview \textbf{improved understanding of data flow and order of execution}. Especially in the domain of data engineering, the combination of being able to know how the underlying data that is manipulated by a program is changed as well as in what order source code is executed is important. For example, S37 wrote, \say{(...) since we have a syntax that very well shows the actual flow of the pipeline (via the block -$>$ block -$>$ ... syntax), it also easily understandable what blocks are executed in which order.}

\begin{boxsummary}
\textbf{Summary:} A data pipeline overview can be separated from implementation details in source code. The enforced structure of Jayvee means this overview always exists, while this is not true for Python/Pandas.

\begin{itemize}
    \item\textit{Ignoring not needed code} improves speed and understanding. However, additional navigation can mean the effect becomes negative if reading details are required.
    \item\textit{The existing overview} improves understanding of data flow and order of execution.
\end{itemize}
\end{boxsummary}

\subsubsection{PL2: Code Structure}
\label{theme:pl2}
Code structure refers to both the way source code is structured, as well as the amount of structure that is enforced by the language. The most significant difference in the way code is structured is the use of the pipes and filters architecture, with connected blocks in Jayvee compared to the script-style implementation in Python/Pandas.

Regarding the amount of enforced structure, Jayvee is much stricter than Python/Pandas. As a general-purpose programming language, Python must allow for more flexibility to enable developers to implement a wider range of programs. In contrast, as a domain-specific language, Jayvee can enforce a structure that is very close to the domain of data pipelines.

This \textbf{consistently enforced structure} enables most survey participants to understand Jayvee better, e.g., S29: \say{Big difference is the structure which Jayvee kind of enforces and developer can easily recognize.} The improved recognition of the structure due to how consistently it is applied is a major element of the positive effect on understanding. 

The \textbf{use of blocks to structure data pipeline code} is highlighted as a positive influence on pipeline understanding, especially for non-professional programmers. For example, S8 likens the experience of using blocks to using LEGO: \say{The best part in Jayvee is block type coding, it is similar to LEGO and you can easily remember, read and write your code.}

Of course, a similar code structure can be achieved using Python with functions or classes but the increased flexibility means that it is not enforced and often not done as S26 points out: \say{The concept of blocks: You can manually create this in Python, but hardly anybody will do this.}

Lastly, \textbf{the encapsulation of related code} is described by participants as making it easier to understand the data pipeline. S44 writes: \say{Jayvee is much easier to understand because every step is divided into blocks the block types are very easy to understand. A single operation is performed in one block, which makes it easy to comprehend.} Importantly, encapsulated code must be sliced so that only a single operation is done in one unit, or participants consider it a detractor for understanding.

\begin{boxsummary}
\textbf{Summary:} Code structure refers to the way source code is organized. Different languages enforce a more or less consistent structure.

\begin{itemize}
    \item\textit{Stricter enforcement of structure} improves understanding and increases learning effects from other data pipelines.
    \item\textit{Consistent structure} allows readers to quickly find expected elements, such as the data pipeline overview.
    \item\textit{Using blocks} is a positive influence on pipeline understanding and aligns with the mental model of data pipelines.
    \item\textit{Encapsulation of related code} makes it easier to understand data pipelines, as long as a single operation is performed in each section.
\end{itemize}
\end{boxsummary}

\subsubsection{PL3: Transparency}
\label{theme:pl3}
Transparency relates to how deeply participants can understand the operations performed in the data pipeline by just reading the source code. Differences can come from how visible implementation details are, depending on the level of abstraction a language aims for. Additionally, how much functionality can be expressed in few lines of code (which we call density of functionality) affects transparency in the sense that with high density of functionality less low-level operations are expressed in source code.

Python/Pandas was identified as having a much higher \textbf{density of functionality} than Jayvee. Regarding the effects, participants had mixed impressions. On one side, being able to express a lot of logic in a few lines of code makes each individual line of code harder to understand, potentially decreasing correctness as S30 explains: \say{Python makes it possible to have a lot of functionality in just a few lines, which can make it hard to read if you have not written it yourself.}

The tradeoff is that pipeline models in a less expressive language must consist of more source code which is slower to read. S0 mentions this concern: \say{Especially in a large pipeline a file might get really big because of all the definitions (especially unnecessary empty block definitions).}. However, because the data pipeline models in our experiment were comparatively small, the majority of participants did not describe this problem.

One way to achieve a high density of functionality is to implement a high \textbf{degree of automatic decisions} and many operations in one unit of code. As an example, loading data with \lstinline{read_csv()} can use various sources and automatically chooses structure and data types based on the underlying data that cannot be inferred from the source code alone. Additionally, the structure of the output can potentially change without any change in the source code if the input data changes. 

Increased automation by grouping many operations in one unit of code makes data pipelines harder to understand and decreases correctness. Often, library methods of Pandas are singled out by participants for this kind of complexity, with S0 remarking: \say{Difficult: The methods sometimes do many things at once (example: load to a sqlite file and automatically choose data types).} S26 describes a similar experience: \say{Functions like \lstinline{pd.read_csv} are hard to understand, as they can read a DF from so many sources (in Jayvee you have one datasource specified).}

Instead of increased automation, the \textbf{inability to see all implementation details} was identified as a negative effect on the ability to understand the data pipeline by participants. This effect was mostly found in Jayvee, with examples including the \lstinline{TableTransformer} block that takes input columns and output columns as properties, for which participants were unsure if it keeps or removes the input columns.

\begin{boxsummary}
\textbf{Summary:} Transparency relates to how well participants can understand every operation performed in a data pipeline based on the source code alone.

\begin{itemize}
    \item\textit{High density of functionality}, many operations per line of code, is a challenge to understanding for small data pipelines. However, reading larger data pipelines will be slow and potentially error-prone with lower density of functionality.
    \item\textit{Increased automation} makes data pipelines harder to understand and decreases comprehension correctness.
    \item\textit{Hidden implementation details} can negatively affect the understanding of data pipelines.
\end{itemize}
\end{boxsummary}

\subsubsection{PL4: Amount of Options}
\label{theme:pl4}
A common theme in the survey responses was the large number of options to implement functionality in Python/Pandas and the comparatively few options in Jayvee. For example, to download a CSV file, Python programmers could use the standard library with \lstinline{urllib} or use Pandas \lstinline{read_csv()} with nearly equivalent outcomes. DSLs can focus on a few core features and only provide one solutions for these.

The effect of \textbf{many competing options} was described as a detriment to understanding by participants such as S49: \say{In Python, there are many varieties and different options, libraries etc, it is harder for non-experienced to grasp the essence.} As they describe, these challenges impact mostly non-professional programmers or programmers unfamiliar with the language itself.

\textbf{External libraries} exacerbate this effect, adding additional ways to solve problems with potentially multiple libraries that solve the same set of problems. Moreover, every library has its own mental model of the problem space with their own glossary, code styles and documentation. S0 writes: \say{In Jayvee everything (all blocks) are from the same source, while in Python there are many libraries with different method styles and documentation.}

External libraries also evolve independently of the main language and each other. This means developers must keep up with changes from different sources to keep their understanding of source code up-to-date, or risk interpreting new library code wrongly.

Despite the challenges that external libraries introduce, their availability has obvious upsides, e.g., less work to implement common functionality. Managing the scope of language features and how external libraries are used is therefore a tradeoff that depends on the experience level of the main users of the language.

\begin{boxsummary}
\textbf{Summary:} The amount of options to implement the same functionality varies greatly between languages, with GPLs having to be more flexible than DSLs. External libraries add additional approaches.

\begin{itemize}
    \item\textit{Many competing options} to solving the same problem are a challenge to understanding data pipelines, mainly for less experienced readers.
    \item\textit{External libraries} increase the amount of available options and have different mental models and glossaries. However, aside from their negative effect on understanding, external libraries reduce required work to implement data pipelines.
\end{itemize}
\end{boxsummary}

\subsubsection{PL5: Syntax}
\label{theme:pl5}
Participants sometimes commented on the syntax differences of the languages as reasons for their performance. Both languages were described as \textbf{human-readable}, sometimes as being like English text or pseudocode. Human-like language syntax was generally linked to making it easier to understand the data pipeline, e.g., by S31: \say{Jayvee has a very human-like language, almost like pseudocode which can be immediately understood even by non programmers in my opinion as long as they have a basic theoretic knowledge about pipelines.}

While Python is well known for its closeness to pseudocode, Jayvee uses considerably more special characters and an uncommon structure. We attribute the positive comments on Jayvee's human-like syntax largely to the use of a glossary that is close to the problem domain, e.g., the use of domain entities such as \lstinline{pipeline} as part of the syntax. Reusing a glossary that is familiar to domain experts allows them to more easily understand the meaning of data pipeline code.

In contrast, encountering \textbf{unfamiliar syntax} is described as a challenge to understanding data pipelines from code. This was mostly an issue for participants solving tasks in Jayvee as they had less previous experience with the language. However, some participants described similar problems with the syntax used by libraries in Python, for example, Pandas creating new columns in a Dataframe with an assignment operator instead of a function call.

\begin{boxsummary}
\textbf{Summary:} Language syntax is discussed by participants, but largely in regard to personal preference for more familiar languages like Python.

\begin{itemize}
    \item\textit{Human-readable} syntax makes it easy to understand a data pipeline. Both Python and Jayvee are described as human-readable languages.
    \item\textit{Unfamiliar syntax} has a negative effect on understanding. New languages and unfamiliar external libraries can introduce this effect.
\end{itemize}
\end{boxsummary}

\subsubsection{PL6: Language Elements}
\label{theme:pl6}
Language elements have a large influence on understanding of data pipeline code. GPLs such as Python must by necessity also provide general-purpose language elements, such as classes or functions, that can be used to build systems for any use case. In contrast, DSLs can express domain concepts such as pipelines, blocks and pipes, or value types directly as language elements.

The \textbf{use of domain-specific language elements} is described as making it easier to understand the data pipeline by participants. The explicit blocks and pipes structure that is enforced by Jayvee aligns closely with how users visualize data pipelines. Readers can then directly build their mental model of the data pipeline from the similar representation in the source code.

Other language elements negatively impacted understanding with some participants mentioning that \textbf{Jayvee language elements are unusual} and need to be learned (in contrast to Pythons language elements that are largely known from other GPLs).

An example are value types based on constraints, as S51 points out: \say{I found the Jayvee code structure a bit difficult to understand, mostly the constraints and value type.} A possible explanation could be that value types and constraints align less obviously than blocks and pipes with the visual model of a data pipeline.

For Python, the \textbf{use of advanced programming concepts} was mentioned as a problem participants faced understanding the experiment tasks. Concrete examples are described by S12: \say{Some functions like lambda, list comprehension and implicit operations are not intuitive and require documentation and comments to understand.} Advanced programming elements have to be used carefully and sparingly if the goal is to create a data pipeline that can be understood by relative junior programmers.

\begin{boxsummary}
\textbf{Summary:} Python must provide general-purpose language elements such as classes and functions, while DSLs can introduce domain concepts such as pipes and blocks.

\begin{itemize}
    \item\textit{Using blocks as domain-specific language elements} improves pipeline understanding and is intuitive because it aligns with the visual model of a data pipeline.
    \item\textit{Unusual language elements} such as value types based on constraints are a challenge to pipeline understanding.
    \item\textit{Advanced programming concepts} like lambdas or list comprehension make pipeline understanding harder, especially for programmers without previous experience in the language.
\end{itemize}
\end{boxsummary}

\subsubsection{HU1: Required Experience}
\label{theme:hu1}
Understanding data pipeline code is influenced by the previous experience of the reader. Depending on the tool used to implement the data pipeline, more or less experience might be needed. Further, the type of experience also matters. Subject-matter experts are often experts in the data they are working with, but might not have extensive software engineering experience.

The \textbf{need for previous experience with programming} to understand Python/Pandas code is mentioned by multiple participants in their surveys. As a GPL, Python must have many features and allow for a maximum amount of flexibility, which makes it inherently complex. Furthermore, more knowledge of programming is involved because the concepts expressed in the language cannot be domain-specific but have to be generic (e.g., classes and functions). S34 expresses the difference: \say{I think the difference might have mostly to do with how much experience one has in programming; I think that Python might require quite some knowledge to get used to, while Jayvee is a bit easier to understand even as a person with not much programming experience.}

\textbf{The more flexible a language is, the more experience and discipline is needed} to stick to good practices and write code that is easy to understand. With the ease of writing script-style Python code, it is not uncommon for developers to implement prototypes in Python that later on get promoted to production code without a rewrite, creating hard to understand data pipelines.

\begin{boxsummary}
\textbf{Summary:} Required experience refers to the amount of experience required to understand a data pipeline from source code. For reading source code, the main required experience is previous programming.

\begin{itemize}
    \item\textit{Previous experience with programming} is needed to understand Python because of the use of generic programming concepts. In contrast, Jayvee is easier to understand for non-programmers because it is using domain-specific concepts.
    \item\textit{More flexibility} means more experience is needed to follow good habits and make code easily readable.
\end{itemize}
\end{boxsummary}

\subsubsection{HU2: Applicable Experience}
\label{theme:hu2}

\textbf{How closely a language aligns with the mental model of data pipelines} is important to reuse experience outside of software engineering. Participants describe Jayvee's blocks and pipes structure as intuitive because it mirrors how they think about data pipelines. This positively affects understanding, e.g., S35 explains why Jayvee pipelines are easy to understand: \say{Jayvee code steps are directly mapped to the data engineering pipeline lifecycle.}

However, the close match to the mental model must be carefully maintained; otherwise it can lead to confusions. One such mismatch were the interpretation blocks in Jayvee (such as the \lstinline{TextFileInterpreter}) to convert binary data to text data. Participants were confused about what the interpretation blocks did because the level of abstraction was lower than what they expected.

A special case of applicable experience is building up knowledge from previous experience with the same tool. \textbf{High flexibility means even similar pipelines can look very different}. A challenge with the low enforced structure of Python/Pandas is that learning effects from creating or reading other data pipelines are reduced. S29 summarizes the challenge as \say{No structure, every pipeline is a new pipeline.} This effect is worsened by the amount of different libraries that can be used to solve common problems, meaning experienced in one library does not necessarily apply to data pipelines that use a different library.

\begin{boxsummary}
\textbf{Summary:} Being able to reuse experience from other sources, such as working with spreadsheets, means data pipelines can be understood by a wider range of readers. Often, subject-matter experts might lack programming experience but have previous domain experience.

\begin{itemize}
    \item\textit{Alignment of code to the mental model of data pipelines} improves understanding, even without programming experience. However, creating the expected abstraction level is important or readers are confused.
    \item\textit{Learning effects are reduced} when similar pipelines can look different in source code due to high flexibility.
\end{itemize}
\end{boxsummary}

\subsubsection{HU3: Naming}
\label{theme:hu3}

\textbf{Good names improve understanding}, especially for non-professionals. However, as Phil Karlton said \say{There are only two hard things in Computer Science: cache invalidation and naming things.} \footnote{https://martinfowler.com/bliki/TwoHardThings.html}

Generally, participants describe names in Jayvee as easy to understand, probably because they are close to the terminology of the domain of data pipelines. In contrast, survey answers mention Python and Pandas as having inconsistent and sometimes confusing naming, potentially because of the generality required by being a GPL and due to the use of external libraries with an inconsistent glossary.

Well named processing steps, both for language elements and user-defined names, have multiple positive effects. Speed is improved by being able to skim source code and clear names make it easier to understand the data pipeline as a whole, S18 writes: \say{When the blocks are named clearly everything can be seen on one quick view. That makes the pipeline easier to understand.}

Good names must \textbf{follow a consistent approach}, which in turn improves understanding. This is a challenge for a GPL like Python because much of the domain-specific functionality comes from external libraries such as Pandas that have different glossaries and approaches to capturing the domain.

Lastly, under the assumption that names are chosen well, the \textbf{quantity of naming opportunities} is important as well, with a higher quantity of names making it easier to understand a data pipeline. Script-style data pipeline implementation give few opportunities for good naming of steps, meaning developers must resort to comments if they want to communicate reasoning. Due to named blocks, Jayvee provides more naming opportunities, both for language elements and user provided names that explain the intent behind the use of a block.

\begin{boxsummary}
\textbf{Summary:} Naming of elements in a pipeline has a major effect on how easy the resulting source code is to understand.

\begin{itemize}
    \item\textit{Good names} improve understanding by allowing readers to skim the source code and get an overview of the whole pipeline.
    \item\textit{Consistent naming} has a positive effect on understanding. External libraries with their own glossary can make naming less consistent.
    \item\textit{The quantity of human-provided names} is important to communicate intend, with a positive effect on understanding if the names are chosen well.
\end{itemize}
\end{boxsummary}

%% file: discussion.tex
\section{Discussion}
\label{sec:discussion}
Based on the results, a DSL based on a pipes and filters structure can be a valuable tool to build data pipelines with subject-matter experts. Participants with a non-professional programmer background can understand data pipeline source code more correctly, but not faster or more easily.

A possible explanation for the similar speed is that  the participants had considerably more previous experience with Python/Pandas than with Jayvee, which likely influenced how fast they were able to understand the data pipelines in favor of Python/Pandas. This will not be an uncommon situation however, because a new DSL always presents a learning challenge, while many practitioners might already have worked with Python and Pandas. However, the fact that participants were still able to complete the tasks with Jayvee in a similar time indicates that learning a new DSL can be done in limited time and provide other benefits like improved correctness, even for non-professional programmers. 

Additionally, Jayvee is considerably more verbose than Python/Pandas, and therefore took participants longer to read before they could solve the tasks. In the context of open data, the tasks were representative of real-life challenges and based on real open data sets. Most open data sets are small, mostly under 10 MB and published in tabular formats such as CSV \citep{Umbrich2015-vn, Mitlohner2016-qq}. However, for larger scale data pipelines, e.g. in industrial settings a more expressive syntax is needed. For these situations, we expect that the difference in speed for program understanding would increase in favor of Python/Pandas due to Jayvee's verbosity and structure.
  
Similarly, more complex tasks could require functionality outside the limited feature set of Jayvee. In previous studies, we have found that in these situations perceived implementation difficulty increases sharply, and it stands to reason that program understanding would decrease as well \citep{Heltweg2025-uc}.

During the experiment, both Jayvee and Python/Pandas source code was displayed as text, without syntax highlighting or the use of an IDE. We chose to not provide an IDE because the maturity of tool support for Python/Pandas and Jayvee differs significantly and would have introduced a confounding factor. In similar work, replication studies of experiments with the addition of IDE support have shown that correctness improves for all treatments, but the relative differences between them remain consistent \citep{Kosar2018-ck}. Therefore, we expect that the results of our experiment would not change significantly with the addition of IDE support.

The code structure of the Python/Pandas data pipelines might have an effect on the results. We chose to use script-style implementations in Python with Pandas, as they are common in practice for smaller data pipelines As discussed in \autoref{theme:pl2}, classes and functions can be used in Python to create a structure similar to Jayvee which would reduce the effects of using a DSL.

With regard to task design, we chose to focus on comprehension tasks of data pipeline structure as a first step. Alternative task goals, such as locating errors or predicting the output of a data pipeline could be used in future work. We consider the comprehension of data pipeline structure as a necessary prerequisite for these tasks. From the qualitative feedback, we expect that the results would be similar for correctness, with Jayvee being more verbose and prescriptive with less functionality. Especially the exact structure of data pipeline output was often unclear to participants due to the automated Dataframe structure creation when loading a data set with Pandas.

Of course, program understanding is only one part of the software development process and other tasks such as extending existing programs or code creation would likely show very different results. We expect implementations in Jayvee to be slower due to the increased verbosity and more strict structure, but additional studies are needed to verify these assumptions.

\subsection{Learnings for Language Designers}
Multiple design decisions are contributing factors to the improved performance and can provide guidelines for future developers of DSLs.

Representing a data pipeline with blocks and pipes as first class language elements seems to be a good choice. It is described as intuitive and clear, especially because it clearly aligns with the mental model of data pipelines as the reader visualizes them.

A data pipeline overview that is represented directly in the syntax of the source code and separated from the implementation details is consistently highlighted as an important positive influence. In addition, the strongly enforced structure of a data pipeline program means readers can quickly orient themselves in the source code and learn with every pipeline they read.

The effect of well-named language elements was considerable, indicating that names are a major influence on data pipeline understanding and especially to provide context to implementation decisions. Consequently, language designers should pay attention to not only using a consistent glossary to name language elements, but also to providing opportunities for developers to use many descriptive names. As an example, by encapsulating functionality into named blocks, data pipelines implemented in Jayvee have a greater minimum amount of named elements than script-style implementations in Python/Pandas. Because this structure is strict, even non-professional programmers are guided to describe the steps they implement in any given pipeline.

Regarding complexity, providing multiple options that achieve the same goal, both in syntax as well in approaches to solve a problem, has been discussed as a barrier to understanding by participants. Because of this, introducing additional syntax or syntactic sugar to make one specific use-case easier should always be seen as a tradeoff between the expressiveness of the language versus the added complexity.

%% file: limitations.tex
\section{Limitations}
\label{sec:limitations}
As a mixed-method study, multiple sets of limitations are potentially relevant to correctly evaluate the results. We evaluate limitations and ways to mitigate them in regard to the quantitative data from the \autoref{subsec:controlled-experiment} and the survey questions, based on threats to validity described in \citet{Wohlin2012-ze}. Trustworthiness criteria according to \citet{Guba1981-bb} are used for the follow-up qualitative work with answers from the descriptive survey (\autoref{subsec:descriptive-survey}).

While we present more than one set of limitations in this chapter, it is important to highlight that the mixed-method approach of this study (with data- and method-triangulation) allows the individual methods to partially make up for the weaknesses of the other. This means the overall research design contributes as a mitigating factor for some of the discussed limitations.

\subsection{Threats to Validity}

We describe potential threads to validity according to the framework presented in \citet{Wohlin2012-ze}.

\subsubsection*{Conclusion Validity}
Threats to conclusion validity are challenges to understanding the correct relationships between the treatment and results of an experiment.

The DSL that was investigated as treatment is in large parts designed and implemented by the authors of this study, therefor bias and searching for positive results is a clear threat to conclusion validity. In an attempt to reduce its impact, we defined the research design as well as hypotheses to analyze ahead of data collection, based on indicators found in previous work \citep{Heltweg2025-uc} and used standard research designs and statistical tests. Additionally, we reported effect sizes and the results of all hypotheses tests, including ones without statistically significant results such as time spent on task.
During data collection, we followed an experiment procedure document to reduce the introduction of individual bias when guiding participants through the experiment. In addition, participants purely interacted with an automated experiment tool that implemented the treatment and took measurements impartially without interaction by the researchers.
Nonetheless, subconscious bias remains as a threat to conclusion validity. Therefore, we have shared the experiment tool\footref{fn:replication-package} to allow for thorough review and independent replication.

Normally, the heterogeneity of students as participants also provides a challenge. However, the use of a crossover experiment design mitigates this concern because they measure differences in comparison to the participants' average and not between participant groups \citep{Vegas2016-dr}.

\subsubsection*{Internal Validity}
Internal validity describes the extent to which influences outside the control of the researcher, apart from the treatment, influence the results of the experiment.

If the tools or tasks used for the experiment were of low quality, they could introduce external factors to the results. In order to reduce these influences, we tested the tool and task implementations in multiple sandbox tests with other researchers and in pilot experiments with individual students from earlier semesters and adjusted them based on feedback, as suggested by \citet{Ko2015-ti}.

Before the experiment runs, one of two researchers explained the experiment procedure to participants and answered questions. Differences in communication style could introduce a threat to internal validity. We mitigated this by preparing an experiment procedure document that was followed by both researchers. In addition, due to the crossover design, every experiment cohort that was instructed by one researcher completed tasks with both treatments and the experiment results depend on the delta in their individual performance, not between groups. Nonetheless, the use of multiple researchers to instruct the participants could have influenced the results between groups.

By selecting volunteers out of a class of students, the results may be influenced if participants think positive responses in regard to Jayvee would have a positive influence on their grade. We therefor clearly communicated to students that data would be anonymized and participation or performance in the experiment would have no effect on their grade.

The differences in previous experience with Jayvee compared to Python/Pandas also introduces a threat to internal validity. We mitigated this by introducing Jayvee with two lectures and at least one practical exercise before the experiment. We also collected and reported the previous experience of participants with both languages to allow for a better contextualizing of the results at the start and end of the semester, but not directly before the experiment. It is likely that the differences in previous experience with the languages influenced the results, especially regarding speed and perceived difficulty. However, we consider the results interesting, because due to its popularity, data practitioners often have previous experience in Python/Pandas and not in new DSLs. We consider our study as a first step to establish initial insights. In further work, replication studies with more balanced previous experience would be needed to confirm the results.

Crossover designs introduce the threat of carryover and familiarization effects, in which the administration of one treatment might influence others. It must be explicitly discussed as a threat to internal validity according to \citet{Vegas2016-dr}. We minimized carryover during the experiment design time in multiple ways. First, by randomly assigning participants to different treatment orders. Second, to reduce the effect of increasing familiarity with the experiment tool itself influencing later task performance, we added an initial task using pseudocode and placeholder step names before applying the real treatments. Lastly, we added a stage of hidden source code, so participants could read the available steps in the pipeline first to reduce the effect of recognizing some steps from the previous task.

Regardless of these measurements, we must recognize that carryover could still be an influencing factor on the results and aim for future replication with between-subject designs.

\vspace{1em}
\subsubsection*{Construct Validity}

Construct validity is concerned with the appropriateness of the experiment construct to measure the underlying concept or theory and the ability to generalize the result of the experiment to it.

The dependent variables in the experiment were clearly defined and measured programmatically. Time and correctness are the most common measures used in bottom-up code comprehension experiments \citep{Wyrich2023-ji}. The concrete definition of correctness for a data pipeline that we used is not previously validated; however, we consider it appropriate because it covers the correct understanding of both selection and the order of steps.

Because only one measurement was taken for each construct, mono-method bias is a concern for the controlled experiment part of this study. This limitation is mitigated by the fact that additional insights about the underlying concepts are drawn from qualitative data as part of the mixed-method design. Nonetheless, additional experiments with more measurements should be done in future work to strengthen the quantitative results.

\vspace{1em}
\subsubsection*{External Validity}

External validity is the ability to generalize the results, e.g., to an industry context.

We chose Masters level students as proxies for a population of subject-matter experts working with data in industry, that are non-professional programmers. When drawing conclusions from the results of this study, it is important to contextualize them with this limited population in mind \citep{Falessi2018-so}. Using students allows us to gather more data points, establish a trend and prepare future studies with practitioners \citep{Tichy2000-nq}. Additional experiments, replicating the same setup, with real subject-matter experts from industry would be needed, but we expect the results to generalize well. Other populations, such as professional programmers from industry, would very likely encounter different challenges and the results of this study should not be taken as indication for their experience.

Because we allowed students to voluntarily opt in to the experiment, only 57 of the 98 students that completed the course participated. We consider this number to be high enough to be representative of the population, however it is possible that less invested students did choose to skip the experiment.

\vspace{1em}
\subsection{Trustworthiness criteria}
For the descriptive survey, we use the trustworthiness criteria of credibility, transferability, dependability, and confirmability \citep{Guba1981-bb}.

\subsubsection*{Credibility}
Our goal was to establish credibility, how well the findings represent the real effects, with various types of triangulation in the mixed-methods research design \citep{Thurmond2001-zx}. By combining the quantitative data from a controlled experiment with the qualitative data of the descriptive surveys, we establish method and data triangulation. In addition, large parts of the qualitative data were coded by multiple researchers as a form of investigator triangulation.

The opt-in, voluntary nature of the experiment introduces a potential bias in the participant selection for more motivated students. We mitigated this effect but clearly stating that participation would have no effect on course grades, both verbally and in the experiment handout we provided to participants.

\subsubsection*{Transferability}

Transferability, how well the results apply to other contexts, has to be discussed from multiple angles.
First, the use of students as participants is problematic when attempting to generalize to professionals in industry, additional context is provided in the discussion regarding the external validity of the experiment that also applies to the qualitative part of the study.

Second, the responses of participants must be seen in the context of one specific DSL, Jayvee, and might not transfer to other DSLs. The descriptions of themes should be seen under this aspect, and additional research with different DSLs is needed to make sure the findings transfer to other languages.

Lastly, the data pipelines that participants had to understand during the experiment were relatively small (but based on real-world open data sets). How well the results transfer to larger scale data pipelines is unclear. When appropriate, we discussed the potential trade-offs regarding small and large data pipelines in the descriptions of the themes (e.g., regarding density of functionality).

To increase transferability, we provided thick descriptions of the themes and extensive quotes from participants in support (as well as an additional, extended description of the themes \footref{fn:replication-package}). Future researchers can use this additional context to evaluate the research results in additional contexts.

\vspace{1em}
\subsubsection*{Dependability}

For dependability, making sure the findings are consistent and can be repeated, we reported the research design in detail and provided as much data as possible. In addition, the complete survey question export and code used to analyze the data is available.

\vspace{1em}
\subsubsection*{Confirmability}

Confirmability, how well the findings represent the objective reality and are not influenced by researcher bias, is challenged by the involvement of the authors in the implementation of Jayvee. Because this introduces a risk of bias, we took steps to introduce additional data and method triangulation by prefacing the survey with a controlled experiment with automated measurements that is less subjective to researcher bias. Regardless of the mitigations employed, we have to acknowledge our own bias and would welcome replication by neutral parties.
To enable other researchers to confirm our findings, we have established an audit trail by describing the research design in detail and providing as much data used during the analysis as possible. Thick descriptions of the themes and direct quotes from the survey also give additional context to the findings.

%% file: conclusion.tex
\section{Conclusion}
\label{sec:conclusion}
In this mixed-methods study, we have asked two research questions: First, \emph{do data pipelines implemented in Jayvee change bottom-up program structure comprehension compared to Python/Pandas for non-professional programmers regarding speed, correctness and perceived difficulty?} and second, \emph{what reasons exist for effects on bottom-up program comprehension for data pipelines implemented in Jayvee compared to Python/Pandas for non-professional programmers?}

To do so, we have executed a controlled experiment with 57 volunteers students comparing their performance on data pipeline understanding tasks implemented in Jayvee and Python with Pandas. In addition, participants could provide qualitative feedback in a post-experiment survey that we then analyzed using qualitative data analysis.

Based on the experiment data, participants are neither faster, nor consider it easier to understand a data pipeline implemented in Jayvee compared to Python/Pandas (\autoref{fig:experiment-time}, \autoref{tab:experiment-time}). However, participants can understand a data pipeline significantly more correctly (\autoref{fig:experiment-correctness}, \autoref{tab:experiment-correctness}).

Qualitative analysis of participant feedback revealed a variety of possible reasons for these effects, summarized in \autoref{fig:codebook}. Data pipelines in the experiment were based on real-life open data sets, but relatively small and further studies would be needed to verify that these effects generalize to larger and more complex data pipelines.

Predictably, most effects are grounded in the difference between programming languages themselves.
Participants highlight the pipeline overview provided by Jayvee as a major positive influence on understandability.
This overview is enforced due to the more rigid structure of Jayvee programs that make them easier to understand than Python/Pandas scripts.
How deeply participants could understand the data pipeline, the transparency of source code, had mixed effects, with high density of functionality and increasing automation making a pipeline harder to understand but faster to read.
Similarly, the amount of available options, especially with the introduction of external libraries, is a challenge to understandability but reduces the work needed to implement pipelines in the first place.
Unfamiliar syntax was an additional problem for some participants, even if both Jayvee and Python were described as human-like languages.
Lastly, provided language elements are a factor in the different outcomes because, as a domain-specific language, Jayvee could include language elements that were intuitive to understand in a data pipeline context while some participants struggled with advanced programming concepts like lambdas in Python.

In addition to the effects of the programming languages themselves, we also identified several human effects.
First, the previous experience required to understand data pipelines from source code differs between the approaches. Participants identify previous programming experience as a necessary precursor to understanding data pipelines written in Python/Pandas, while they consider pipelines written in Jayvee to be approachable by novices.
Second, the implementation language effects which previous experience is applicable to understanding a data pipeline. If the abstraction level is maintained well, a domain-specific language like Jayvee allows readers to reuse previous experience from data engineering with other tools like visual modeling software.
Finally, depending on the reader, well-chosen, descriptive names have a large influence on how understandable data pipeline source code is. Languages with a wide library ecosystem like Python with Pandas face challenges to keep a consistent glossary between different authors. Additionally, the strict structure of Jayvee with extensive possibilities for user-provided names allowed future readers to infer additional information.

Besides the effects that are often described and have a clear influence, open questions remain. For example, the best abstraction level of a domain-specific language for data pipelines is unclear and might depend on the intended audience. Additionally, a good tradeoff between the reuse of work with a library ecosystem versus the complexity it introduces warrants further studies. Density of functionality shows a similar tradeoff between short to write and expressive code versus harder to understand pipelines. With more research, it might be possible to identify the reasons for the largest negative effects and avoid them in future language design.

In summary, domain-specific languages such as Jayvee have the potential to be more correct in the domain of data pipeline modeling. These effects are especially strong for non-professional programmers, such as subject-matter experts in other domains. A variety of reasons for these effects exists, largely based on the programming language itself or on the type of reader that tries to understand the source code. However, the exact effect of many reasons is still an open question that needs further research to develop a comprehensive theory of domain-specific languages for data pipeline modeling.

In future work, we intend to explore more narrow features of domain-specific languages for data engineering, such as value types or selection syntax for tabular data, with additional controlled experiments.

%% file: acknowledgements.tex
\section*{Acknowledgements}
This research has been partially funded by the German Federal Ministry of Education and Research (BMBF) through grant 01IS17045 Software Campus 2.0 (Friedrich-Alexander-Universität Erlangen-Nürnberg) as part of the Software Campus project ’JValue-OCDE-Case1’. Responsibility for the content of this publication lies with the authors.

%% file: data-availibility-statement.tex
\section*{Data Availability Statement}
\label{sec:data-availability-statement}

The data generated and analyzed during the current study is available on Zenodo at:

\url{https://doi.org/10.5281/zenodo.15574873}.

\noindent For convenience, the full codebook is also hosted at:

\url{https://rhazn.github.io/2025-data-release-program-comprehension-jayvee/}.

%% file: conflict-of-interest.tex

%% file: appendix.tex
\section*{Appendix}
\label{sec:appendix}

\subsection{Task Examples}
\begin{figure}[!ht]
    \input{task-comparison-schema.tex}
    \caption{Comparison of source code to filter and apply a schema to data, shown for task 2 in Jayvee and Python/Pandas.}
    \label{fig:task-comparison-schema}
\end{figure}

\subsection{Extended Result Data}

\begin{figure}[!ht]
    \includegraphics[width=\textwidth]{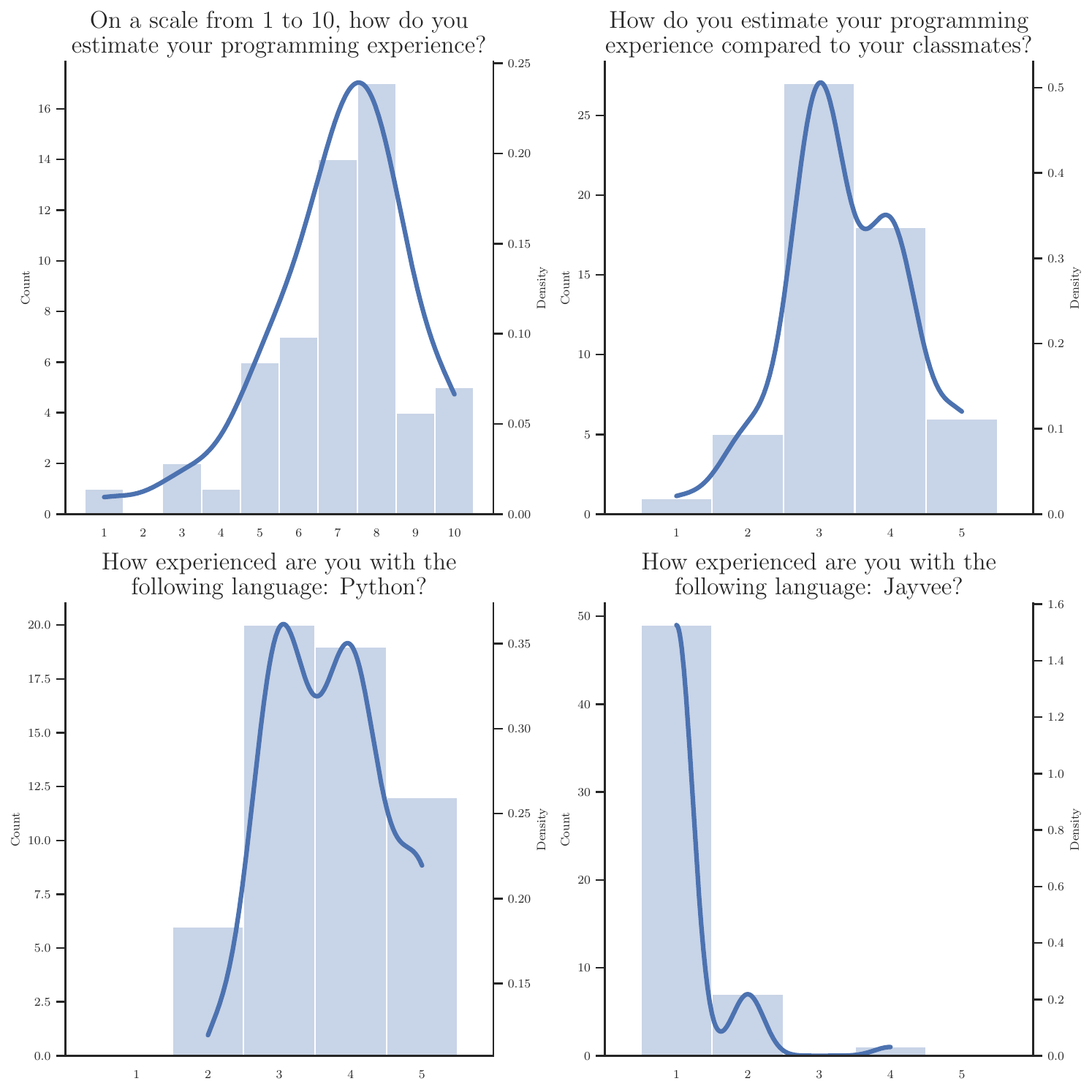}
    \caption{Previous experience of experiment participants.}
    \label{fig:experiment-participants}
\end{figure}

\begin{figure}[!ht]
    \includegraphics[width=\textwidth]{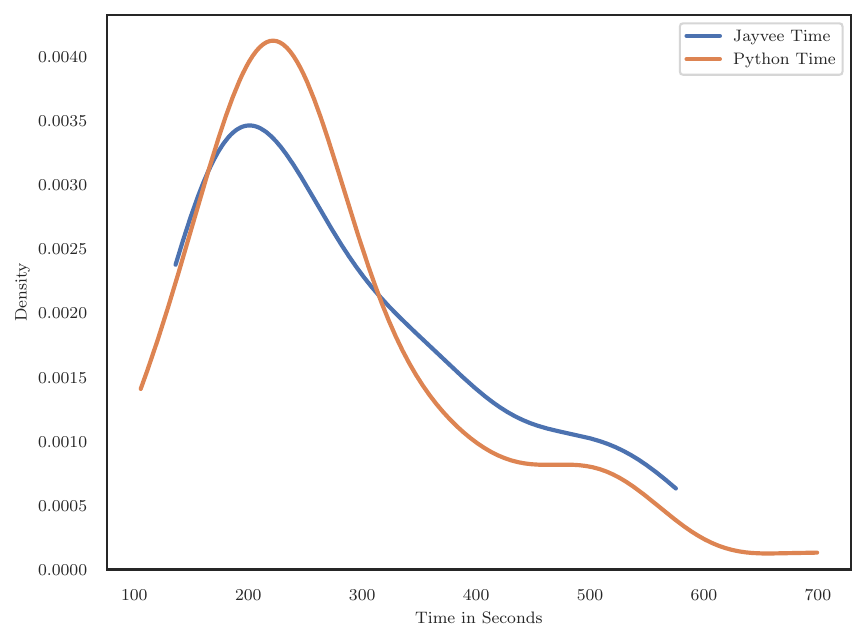}
    \caption{Kernel-density-plot of time on task for Jayvee compared to Python/Pandas.}
    \label{fig:experiment-time}
\end{figure}

\begin{figure}[!ht]
    \includegraphics[width=\textwidth]{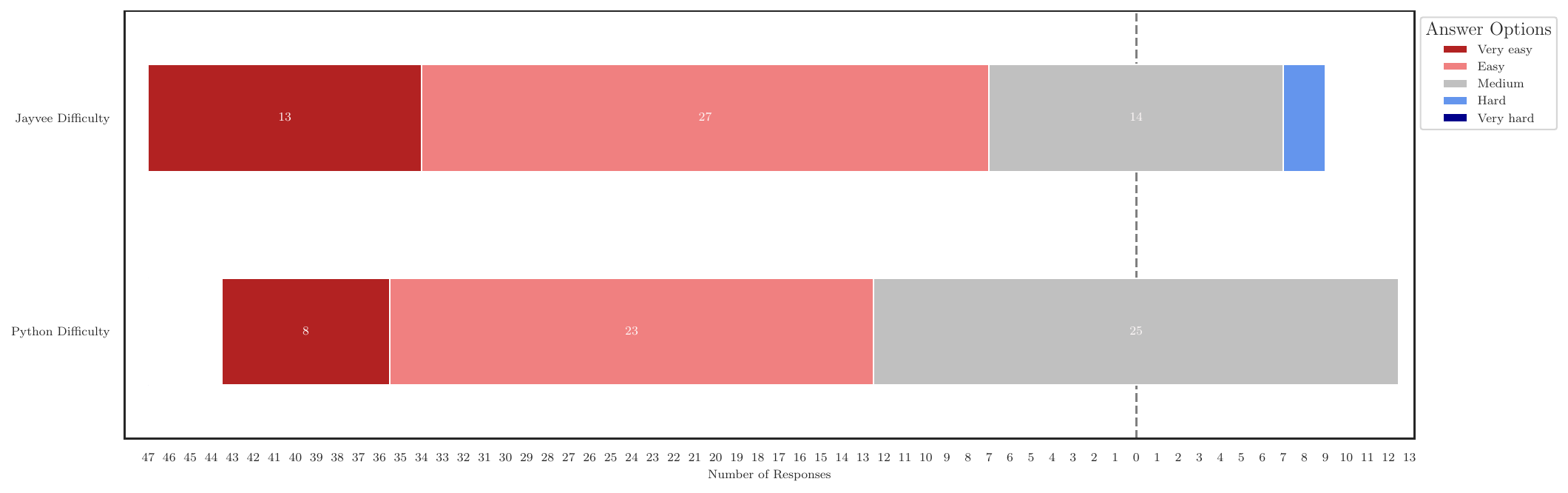}
    \caption{Diverging stacked bar charts according to \citet{Robbins2011-lh} and \citet{Heiberger2014-fs} for perceived difficulty of using Jayvee compared to Python/Pandas.\textsuperscript{*}}
    \tiny\textsuperscript{*} One outlier participant (S25) considered using Jayvee hard (and Python/Pandas easy) due to their lack of previous experience with Jayvee and did not provide more details, writing: \say{(Jayvee) is new so I think it was not easy to understand or read.}
    \label{fig:survey-difficulty}
\end{figure}

%% file: task-comparison-schema.tex
\noindent
  \begin{minipage}[t][23\baselineskip][t]{.55\textwidth}
      \begin{lstlisting}[language=Jayvee]
constraint GeographicCoordinateScale on decimal: value >= -90 and value <= 90;

valuetype GeographicCoordinate oftype decimal {
    constraints: [GeographicCoordinateScale];
}

block ValuetypeValidator oftype TableInterpreter {
  header: true;
  columns:[
    'uuid' oftype text,
    'latitude' oftype GeographicCoordinate,
    'longitude' oftype GeographicCoordinate,
    'bezeichnung' oftype text,
    'traeger_bezeichnung' oftype text,
    'traeger_art' oftype text,
    'website' oftype text,
  ];
}
    \end{lstlisting}
  \end{minipage}
  \hfill
  \begin{minipage}[t][23\baselineskip][t]{.39\textwidth}
    \begin{lstlisting}[language=Python]
data = data[[
  'uuid',
  'latitude',
  'longitude',
  'bezeichnung',
  'traeger_bezeichnung',
  'traeger_art',
  'website',
]]

data = data.astype({
  'uuid': str,
  'latitude': float,
  'longitude': float,
  'bezeichnung': str,
  'traeger_bezeichnung': str,
  'traeger_art': str,
  'website': str,
})

data = data[data['latitude'].apply(lambda input: input >= -90 and input <= 90)]
data = data[data['longitude'].apply(lambda input: input >= -90 and input <= 90)]
    \end{lstlisting}
  \end{minipage}

%% file: arxiv.bbl
\begin{thebibliography}{}

\bibitem[Barišić et~al., 2018]{Barisic2018-kt}
Barišić, A., Amaral, V., and Goulão, M. (2018).
\newblock Usability driven {DSL} development with {USE}-{ME}.
\newblock {\em Computer languages, systems \& structures}, 51:118--157.

\bibitem[Basili and Rombach, 1988]{Basili1988-al}
Basili, V.~R. and Rombach, H.~D. (1988).
\newblock The {TAME} project: towards improvement-oriented software environments.
\newblock {\em IEEE Transactions on Software Engineering}, 14(6):758--773.

\bibitem[Bowen, 2008]{Bowen2008-xh}
Bowen, G.~A. (2008).
\newblock Naturalistic inquiry and the saturation concept: a research note.
\newblock {\em Qualitative research: QR}, 8(1):137--152.

\bibitem[Braun and Clarke, 2012]{Braun2012-cm}
Braun, V. and Clarke, V. (2012).
\newblock Thematic analysis.
\newblock In {\em {APA} handbook of research methods in psychology, Vol 2: Research designs: Quantitative, qualitative, neuropsychological, and biological}, pages 57--71. American Psychological Association, Washington.

\bibitem[Buse et~al., 2011]{Buse2011-ra}
Buse, R. P.~L., Sadowski, C., and Weimer, W. (2011).
\newblock Benefits and barriers of user evaluation in software engineering research.
\newblock In {\em Proceedings of the 2011 ACM international conference on Object oriented programming systems languages and applications}, New York, NY, USA. ACM.

\bibitem[Cingolani et~al., 2015]{Cingolani2015-tr}
Cingolani, P., Sladek, R., and Blanchette, M. (2015).
\newblock {BigDataScript}: a scripting language for data pipelines.
\newblock {\em Bioinformatics (Oxford, England)}, 31(1):10--16.

\bibitem[do~Nascimento et~al., 2012]{Do_Nascimento2012-cm}
do~Nascimento, L.~M., Viana, D.~L., Neto, P. A.~S., Martins, D.~A., Garcia, V.~C., and Meira, S.~R. (2012).
\newblock A systematic mapping study on domain-specific languages.
\newblock In {\em The Seventh International Conference on Software Engineering Advances (ICSEA 2012)}, pages 179--187.

\bibitem[Falessi et~al., 2018]{Falessi2018-so}
Falessi, D., Juristo, N., Wohlin, C., Turhan, B., M{\"u}nch, J., Jedlitschka, A., and Oivo, M. (2018).
\newblock Empirical software engineering experts on the use of students and professionals in experiments.
\newblock {\em Empirical Software Engineering}, 23(1):452--489.

\bibitem[Feigenspan et~al., 2012]{Feigenspan2012-gv}
Feigenspan, J., K{\"a}stner, C., Liebig, J., Apel, S., and Hanenberg, S. (2012).
\newblock Measuring programming experience.
\newblock In {\em 2012 20th {IEEE} International Conference on Program Comprehension ({ICPC})}, pages 73--82. Ieee.

\bibitem[Fonseca et~al., 2020]{Fonseca2020-tv}
Fonseca, N., Paulo~Fernandes, J., Pires, M., and Melo~de Sousa, S. (2020).
\newblock {PACE}: A {DSL}-based approach to manage complex build pipelines.
\newblock In {\em 2020 46th Euromicro Conference on Software Engineering and Advanced Applications (SEAA)}, pages 43--50. IEEE.

\bibitem[Garlan and Shaw, 1993]{Garlan1993-cz}
Garlan, D. and Shaw, M. (1993).
\newblock {AN} {INTRODUCTION} {TO} {SOFTWARE} {ARCHITECTURE}.
\newblock In {\em Advances in Software Engineering and Knowledge Engineering}, volume~2 of {\em Series on Software Engineering and Knowledge Engineering}, pages 1--39. WORLD SCIENTIFIC.

\bibitem[Guba, 1981]{Guba1981-bb}
Guba, E.~G. (1981).
\newblock Criteria for assessing the trustworthiness of naturalistic inquiries.
\newblock {\em ECTJ}, 29(2):75.

\bibitem[Heiberger and Robbins, 2014]{Heiberger2014-fs}
Heiberger, R. and Robbins, N. (2014).
\newblock Design of diverging stacked bar charts for likert scales and other applications.
\newblock {\em Journal of statistical software}, 57:1--32.

\bibitem[Heltweg and Riehle, 2023]{Heltweg2023-ps}
Heltweg, P. and Riehle, D. (2023).
\newblock A systematic analysis of problems in open collaborative data engineering.
\newblock {\em Trans. Soc. Comput.}, 6(3-4):1--30.

\bibitem[Heltweg et~al., 2025]{Heltweg2025-uc}
Heltweg, P., Schwarz, G.-D., Dirk, R., and Felix, Q. (2025).
\newblock An empirical study on the effects of jayvee, a domain-specific language for data engineering, on understanding data pipeline architectures.
\newblock {\em Software: Practice \& Experience}.

\bibitem[Hoffmann et~al., 2022]{Hoffmann2022-ab}
Hoffmann, B., Urquhart, N., Chalmers, K., and Guckert, M. (2022).
\newblock An empirical evaluation of a novel domain-specific language - modelling vehicle routing problems with athos.
\newblock {\em Empirical Software Engineer}, 27(7):180.

\bibitem[Jedlitschka and Pfahl, 2005]{Jedlitschka2005-mu}
Jedlitschka, A. and Pfahl, D. (2005).
\newblock Reporting guidelines for controlled experiments in software engineering.
\newblock In {\em 2005 International Symposium on Empirical Software Engineering, 2005.}, page 10 pp. Ieee.

\bibitem[Johanson and Hasselbring, 2017]{Johanson2017-sb}
Johanson, A.~N. and Hasselbring, W. (2017).
\newblock Effectiveness and efficiency of a domain-specific language for high-performance marine ecosystem simulation: a controlled experiment.
\newblock {\em Empirical Software Engineering}, 22:2206--2236.

\bibitem[Johnson et~al., 2007]{Johnson2007-kr}
Johnson, R.~B., Onwuegbuzie, A.~J., and Turner, L.~A. (2007).
\newblock Toward a definition of mixed methods research.
\newblock {\em Journal of mixed methods research}, 1(2):112--133.

\bibitem[Kerby, 2014]{Kerby2014-xq}
Kerby, D.~S. (2014).
\newblock The simple difference formula: An approach to teaching nonparametric correlation.
\newblock {\em Comprehensive psychology}, 3:11.IT.3.1.

\bibitem[Kitchenham et~al., 2017]{Kitchenham2017-oi}
Kitchenham, B., Madeyski, L., Budgen, D., Keung, J., Brereton, P., Charters, S., Gibbs, S., and Pohthong, A. (2017).
\newblock Robust statistical methods for empirical software engineering.
\newblock {\em Empirical Software Engineering}, 22(2):579--630.

\bibitem[Kitchenham and Pfleeger, 2008]{Kitchenham2008-ki}
Kitchenham, B.~A. and Pfleeger, S.~L. (2008).
\newblock Personal opinion surveys.
\newblock In Shull, F., Singer, J., and Sjøberg, D. I.~K., editors, {\em Guide to Advanced Empirical Software Engineering}, pages 63--92. Springer London, London.

\bibitem[Klanten et~al., 2024]{Klanten2024-pf}
Klanten, K., Hanenberg, S., Gries, S., and Gruhn, V. (2024).
\newblock Readability of domain-specific languages: A controlled experiment comparing (declarative) inference rules with (imperative) java source code in programming language design.
\newblock In {\em Proceedings of the 19th International Conference on Software Technologies}. SCITEPRESS - Science and Technology Publications.

\bibitem[Ko et~al., 2015]{Ko2015-ti}
Ko, A.~J., LaToza, T.~D., and Burnett, M.~M. (2015).
\newblock A practical guide to controlled experiments of software engineering tools with human participants.
\newblock {\em Empirical Software Engineering}, 20(1):110--141.

\bibitem[Kosar et~al., 2018]{Kosar2018-ck}
Kosar, T., Gaberc, S., Carver, J.~C., and Mernik, M. (2018).
\newblock Program comprehension of domain-specific and general-purpose languages: replication of a family of experiments using integrated development environments.
\newblock {\em Empirical Software Engineering}, 23(5):2734--2763.

\bibitem[Kosar et~al., 2012]{Kosar2012-ap}
Kosar, T., Mernik, M., and Carver, J.~C. (2012).
\newblock Program comprehension of domain-specific and general-purpose languages: comparison using a family of experiments.
\newblock {\em Empirical software engineering}, 17:276--304.

\bibitem[Kosar et~al., 2010]{Kosar2010-cp}
Kosar, T., Oliveira, N., Mernik, M., João, M., Pereira, M., Repinåek, M., Cruz, D., and Rangel~Henriques, P. (2010).
\newblock Comparing general-purpose domain specific languages: empirical study.
\newblock {\em Computer Science Information Systems}.

\bibitem[Landis and Koch, 1977]{Landis1977-ik}
Landis, J.~R. and Koch, G.~G. (1977).
\newblock The measurement of observer agreement for categorical data.
\newblock {\em Biometrics}, 33(1):159--174.

\bibitem[Lopes et~al., 2021]{Lopes2021-ol}
Lopes, J., Bernardino, M., Basso, F., and Rodrigues, E. (2021).
\newblock Textual-based {DSL} for conceptual database modeling: A controlled experiment.
\newblock In {\em Anais do XXXVI Simpósio Brasileiro de Banco de Dados (SBBD 2021)}, pages 169--180. Sociedade Brasileira de Computação - SBC.

\bibitem[McGraw and Wong, 1992]{McGraw1992-ks}
McGraw, K.~O. and Wong, S.~P. (1992).
\newblock A common language effect size statistic.
\newblock {\em Psychological bulletin}, 111(2):361--365.

\bibitem[Misale, 2017]{Misale2017-wo}
Misale, C. (2017).
\newblock {\em {PiCo}: A Domain-Specific Language for Data Analytics Pipelines}.
\newblock PhD thesis, University of Torino, Italy.

\bibitem[Mitlohner et~al., 2016]{Mitlohner2016-qq}
Mitlohner, J., Neumaier, S., Umbrich, J., and Polleres, A. (2016).
\newblock Characteristics of open data {CSV} files.
\newblock In {\em 2016 2nd International Conference on Open and Big Data ({OBD})}. IEEE.

\bibitem[Robbins et~al., 2011]{Robbins2011-lh}
Robbins, N.~B., Heiberger, R.~M., and {Others} (2011).
\newblock Plotting likert and other rating scales.
\newblock In {\em Proceedings of the 2011 joint statistical meeting}, volume~1.

\bibitem[Roberto~Minelli and Lanza, 2015]{Roberto-Minelli2015-yo}
Roberto~Minelli, A.~M. and Lanza, M. (2015).
\newblock {I} know what you did last summer.
\newblock In {\em 2015 IEEE 23rd International Conference on Program Comprehension}, volume~3, pages 1--28.

\bibitem[Shapiro and Wilk, 1965]{Shapiro1965-fv}
Shapiro, S.~S. and Wilk, M.~B. (1965).
\newblock An analysis of variance test for normality (complete samples).
\newblock {\em Biometrika}, 52(3/4):591--611.

\bibitem[Shaw and Garlan, 1995]{Shaw1995-hz}
Shaw, M. and Garlan, D. (1995).
\newblock Formulations and formalisms in software architecture.
\newblock In van Leeuwen, J., editor, {\em Computer Science Today: Recent Trends and Developments}, pages 307--323. Springer Berlin Heidelberg, Berlin, Heidelberg.

\bibitem[Thurmond, 2001]{Thurmond2001-zx}
Thurmond, V.~A. (2001).
\newblock The point of triangulation.
\newblock {\em Journal of nursing scholarship: an official publication of Sigma Theta Tau International Honor Society of Nursing / Sigma Theta Tau}, 33(3):253--258.

\bibitem[Tichy, 2000]{Tichy2000-nq}
Tichy, W.~F. (2000).
\newblock Hints for reviewing empirical work in software engineering.
\newblock {\em Empirical Software Engineering}, 5(4):309--312.

\bibitem[Umbrich et~al., 2015]{Umbrich2015-vn}
Umbrich, J., Neumaier, S., and Polleres, A. (2015).
\newblock Quality assessment and evolution of open data portals.
\newblock In {\em 2015 3rd International Conference on Future Internet of Things and Cloud}, pages 404--411. IEEE.

\bibitem[Vallat, 2018]{Vallat2018}
Vallat, R. (2018).
\newblock Pingouin: statistics in python.
\newblock {\em The Journal of Open Source Software}, 3(31):1026.

\bibitem[Vargha and Delaney, 2000]{Vargha2000-zj}
Vargha, A. and Delaney, H.~D. (2000).
\newblock A critique and improvement of the {CL} common language effect size statistics of {McGraw} and wong.
\newblock {\em Journal of educational and behavioral statistics: a quarterly publication sponsored by the American Educational Research Association and the American Statistical Association}, 25(2):101--132.

\bibitem[Vegas et~al., 2016]{Vegas2016-dr}
Vegas, S., Apa, C., and Juristo, N. (2016).
\newblock Crossover designs in software engineering experiments: Benefits and perils.
\newblock {\em IEEE Transactions on Software Engineering}, 42(2):120--135.

\bibitem[Wilcoxon, 1945]{Wilcoxon1945-uf}
Wilcoxon, F. (1945).
\newblock Individual comparisons by ranking methods.
\newblock {\em Biometrics bulletin}, 1(6):80.

\bibitem[Wile, 2004]{Wile2004-rr}
Wile, D. (2004).
\newblock Lessons learned from real {DSL} experiments.
\newblock {\em Science of computer programming}, 51(3):265--290.

\bibitem[Wohlin et~al., 2012]{Wohlin2012-ze}
Wohlin, C., Runeson, P., H{\"o}st, M., Ohlsson, M.~C., Regnell, B., and Wessl{\'e}n, A. (2012).
\newblock {\em Experimentation in Software Engineering}.
\newblock Springer Science + Business Media.

\bibitem[Wyrich et~al., 2023]{Wyrich2023-ji}
Wyrich, M., Bogner, J., and Wagner, S. (2023).
\newblock 40 years of designing code comprehension experiments: A systematic mapping study.
\newblock {\em ACM Comput. Surv.}

\bibitem[Xia et~al., 2018]{Xia2018-hl}
Xia, X., Bao, L., Lo, D., Xing, Z., Hassan, A.~E., and Li, S. (2018).
\newblock Measuring program comprehension: A large-scale field study with professionals.
\newblock {\em IEEE transactions on software engineering}, 44(10):951--976.

\end{thebibliography}
